%% file: Kashibadze_n_en-fin.tex
\documentclass[
 prl,
 11pt,tightenlines,
 final,
 notitlepage,
 oneside,
 twocolumn,
 nobibnotes,
 nofootinbib,
 superscriptaddress,
 noshowpacs,
 centertags,
 amsmath,
 amssymb,
 showkeys,
 preprintnumbers]
{revtex4-2}

\usepackage{graphicx}
\usepackage{dcolumn}
\usepackage{bm}
\newcommand\fdg{\mbox{$.\!\!^\circ$}}%

\begin{document}
\preprint{\texttt{2018AstBu..73..124K}}

%

\keywords{galaxies: kinematics and dynamics---galaxies: distances and
redshifts---galaxies: groups}

\let\svthefootnote\thefootnote

\title{\LARGE{}Surveying the Local Supercluster Plane$^*$}
\let\thefootnote\relax\footnotetext{%
\mbox{\hspace{-1em}$^*$} Table~\ref{table01:Kashibadze_n_en} is available at the CDS via anonymous ftp\linebreak
ftp://cdsarc.u-strasbg.fr/pub/cats/J/other/AstBu/73.\linebreak
124 (130.79.128.5) or via http://vizier.u-strasbg.fr/viz-bin/VizieR?-source=J/other/AstBu/73.124}

\author{\bf{}O.\,G.\,Kashibadze$^{**}$} \let\thefootnote\relax\footnotetext{%
\mbox{\hspace{-0.8em}$^{**}$}\rule{0pt}{4mm}phiruzi@gmail.com}%
\affiliation{Special Astrophysical Observatory, Russian Academy of Sciences,
Nizhnii Arkhyz, 369167 Russia}
\author{\bf{}I.\,D.\,Karachentsev}
\affiliation{Special Astrophysical Observatory, Russian Academy of Sciences,
Nizhnii Arkhyz, 369167 Russia}
\author{\bf{}V.\,E.\,Karachentseva}
\affiliation {Main Astronomical Observatory, National Academy of
Sciences of Ukraine,  Kyiv, 03143 Ukraine}

\addtocounter{footnote}{-2}\let\thefootnote\svthefootnote


\begin{abstract}
We investigate the distribution and velocity field of galaxies
situated in a band of 100 by 20 degrees centered on M\,87 and
oriented along the Local supercluster plane. Our sample amounts
2158 galaxies with radial velocities less than 2000~km\,s$^{-1}$.
Of them, 1119 galaxies (52\%) have distance and peculiar velocity
estimates. About 3/4 of early-type galaxies are concentrated
within the Virgo cluster core, most of the late-type galaxies in
the band locate outside the virial radius. Distribution of
gas-rich dwarfs with $M_{\rm H\,I} > M_*$ looks to be insensitive
to the Virgo cluster presence. Among 50 galaxy groups in the
equatorial supercluster band 6 groups have peculiar velocities
about 500--1000~km\,s$^{-1}$ comparable with virial motions in
rich clusters. The most cryptic case is a flock of nearly 30
galaxies around NGC\,4278 (Coma\,I cloud), moving to us with the
mean peculiar velocity of $-$840~km\,s$^{-1}$. This cloud (or
filament?) resides at a distance of 16.1 Mpc from us and
approximately $5$ Mpc away from the Virgo center. Galaxies around
Virgo cluster exhibit Virgocentric infall with an amplitude of
about $500$~km\,s$^{-1}$. Assuming the spherically symmetric
radial infall, we estimate the radius of the zero-velocity surface
to be $R_0 = (7.0\pm0.3)$ Mpc that yields the total mass of Virgo
cluster to be $(7.4\pm0.9)\times 10^{14} M_{\odot}$ in tight
agreement with its virial mass estimates. We conclude that the
Virgo outskirts does not contain significant amounts of dark mater
beyond its virial core.
\end{abstract}

\maketitle

\section{INTRODUCTION}
The large-scale structure of the Universe---voids, walls, and
filaments---evolve through coherent non-Hubble flow motions with
amplitude of several hundred km~s$^{-1}$ on scales of 10--100~Mpc.
Such a pattern is resulting from N-body simulations in the
standard cosmological $\Lambda$CDM
model~\cite{kly2003:Kashibadze_n_en,sch2007:Kashibadze_n_en,cec2016:Kashibadze_n_en}.
Actually, the Milky Way, together with dozens of other neighboring
galaxies forming the planar structure of the {\it Local Sheet},
moves relative to the cosmic microwave background with a velocity
of 630~km~s$^{-1}$~\cite{kog1993:Kashibadze_n_en}. According to
Tully~et~al.~\cite{tul2008:Kashibadze_n_en,tul2016:Kashibadze_n_en},
this velocity vector is composed of three roughly orthogonal
components: (1) infall of the Local Sheet toward the center of the
nearest Virgo cluster with a velocity of 185~km~s$^{-1}$, (2)
outflow from the expanding Local Void with the velocity of
260~km~s$^{-1}$ and (3) bulk motion toward Hydra--Centaurus
cluster and the Shapley
supercluster~\cite{sca1995:Kashibadze_n_en} with an amplitude of
about 450~km~s$^{-1}$. This complicated background of
multidirectional flows makes it useless to estimate distances of
nearby galaxies via the linear Hubble relation: $V_{\rm LG}=H_0
D$, where $D$ is distance of a neighboring galaxy, $V_{\rm LG}$ is
its radial velocity relative to the Local Group centroid and $H_0$
is the Hubble parameter. Even those models of non-Hubble velocity
fields which consider only the Virgo cluster but not other
components~\cite{kra1986:Kashibadze_n_en,mas2005:Kashibadze_n_en}
are yet too simplified to determine distances of Local
Supercluster (LSC) galaxies from their radial velocities.

Large-area sky surveys in neutral hydrogen (H\,I) 21-centimeter
line---HIPASS~\cite{kor2004:Kashibadze_n_en},
ALFALFA~\cite{hay2011:Kashibadze_n_en},
WSRT-CVn~\cite{kov2009:Kashibadze_n_en}---gave an opportunity of
the wholesale measurements of distances to galaxies with an
accuracy of approximately 20--25\% from
Tully--Fisher~\cite{tul1977:Kashibadze_n_en} relation between the
luminosity of a galaxy and the width of its H\,I line. At this
stage the Tully--Fisher distance estimates are made for several
thousand galaxies with typical velocities of about
6000~km~s$^{-1}$~\cite{tul2016:Kashibadze_n_en}, making the basis
for studying cosmic flows in the scale of about 100--200~Mpc.

\begin{figure*} 
\includegraphics[width=\textwidth]{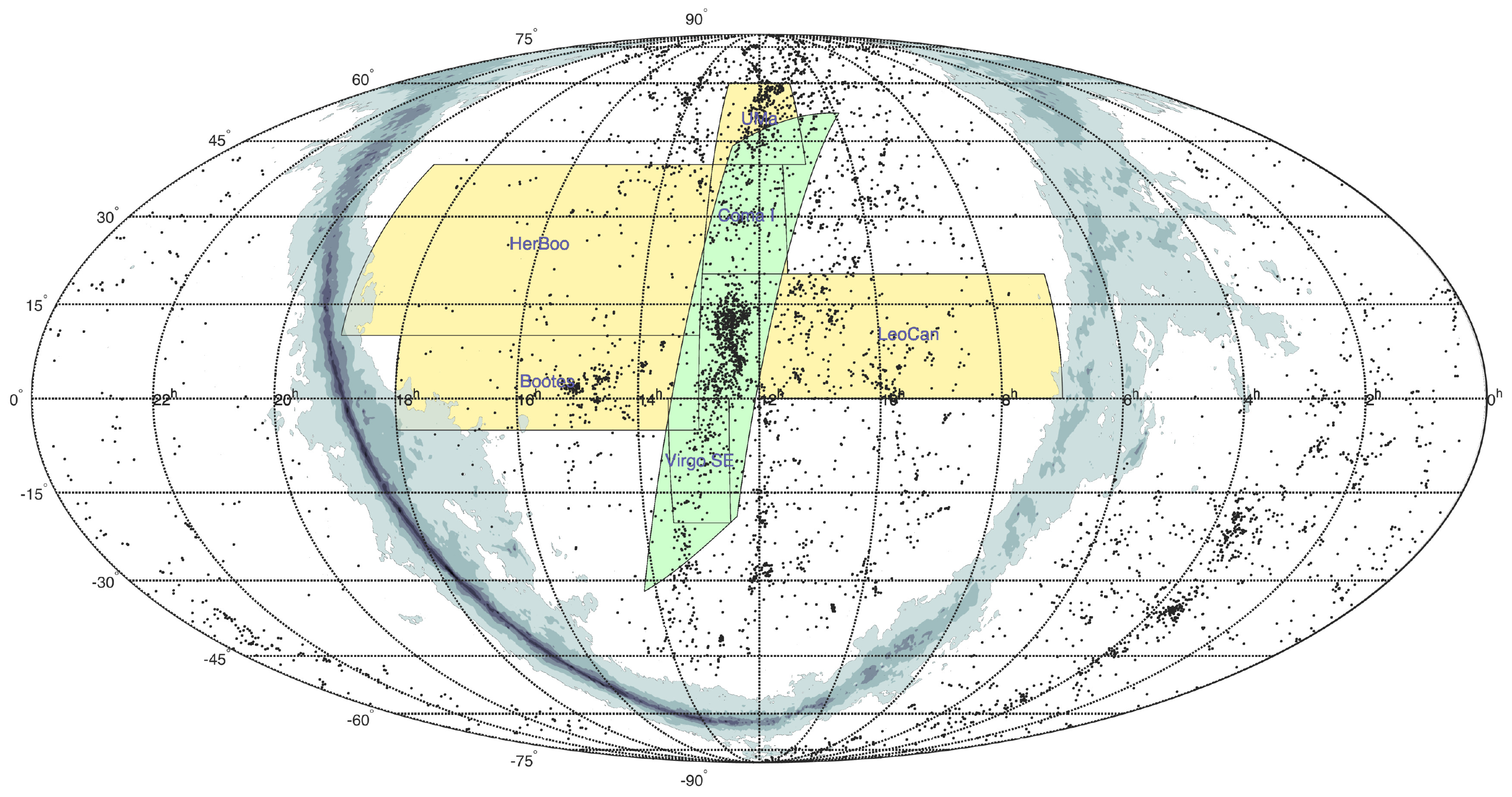}
\caption{The distribution of 5180 galaxies with radial velocities
$V_{\rm LG}< 2000$~km~s$^{-1}$ over the sky in equatorial
coordinates. The Zone of Avoidance near the plane of the Milky Way is
shadowed. The region being considered in this paper is filled with
green. Other sky areas discussed before are marked by yellow. (See
electronic version of the paper for colored figures.)} \label{fig01:Kashibadze_n_en}
\end{figure*}

Inside the Local Supercluster ($D<30$~Mpc) with the Virgo cluster
located at 16.7~Mpc~\cite{mei2007:Kashibadze_n_en} as its core,
there are many spiral and irregular galaxies with measured H\,I
line widths $W_{50}$ but without distance estimates. For many of
them we determined Tully--Fisher distances for the first time,
increasing the density of the peculiar velocity field in the Local
Supercluster more than twice.

In our previous papers we considered motions of galaxies in
several regions along the LSC equator:
Coma\,I~\cite{kar2011:Kashibadze_n_en}, Ursa
Major~\cite{kar2013b:Kashibadze_n_en}, Virgo Southern
Extension~\cite{kar2013a:Kashibadze_n_en}, as well as in broad
bands Leo--Cancer~\cite{kar2015:Kashibadze_n_en},
Bootes~\cite{kar2014a:Kashibadze_n_en} and
Hercules--Bootes~\cite{kar2017a:Kashibadze_n_en} flanking to the
LSC equator. In the last cases, we have obtained new evidence that
the Local Void expands, while galaxies in the vicinity of the
Virgo cluster move toward its center. Beyond that, we had
considered particularly the kinematic situation around the Virgo
and Fornax clusters and the Local
Void~\cite{kar2010:Kashibadze_n_en,nas2011a:Kashibadze_n_en,nas2011b:Kashibadze_n_en}
and estimated the total mass for both clusters, as well as the
density contrast in the Local Void. The map with the discussed
areas distributed over the sky in equatorial coordinates is shown
in Fig.~1 where dots mark 5180 galaxies with radial velocities
$V_{\rm LG}<2000$~km~s$^{-1}$ and the floccose annular band
represents the Zone of Avoidance obscured by the Milky Way.

\section{SAMPLE OF GALAXIES IN THE LSC PLANE}

We used
HyperLEDA\footnote{\protect\url{http://leda.univ-lyon1.fr/}}~\cite{mak2014:Kashibadze_n_en}
and NASA Extragalactic Database
(NED)\footnote{\protect\url{http://ned.ipac.caltech.edu/}} as the main
sources of observational data on LSC galaxies amplifying them by
the most recent estimates of galaxy distances. We have selected
galaxies in the band $\pm10^{\circ}$ by the supergalactic latitude
(SGB) and $\pm50^{\circ}$ by the supergalactic longitude (SGL)
relative to the Virgo cluster center which has been identified
with the radio galaxy M\,87~= NGC\,4486 (${\rm SGL}=102\fdg88$,
${\rm SGB} = -2\fdg35$). The reason for such a choice is that
M\,87 is situated near the barycenter of the hot X-ray emitting
intracluster gas. We have filtered our sample by radial velocities
relative to the Local Group centroid, keeping only galaxies with
$V_{\rm LG}<2000$~km~s$^{-1}$. This condition allows to cover the
essential population of the Virgo cluster and the adjacent
structures avoiding a large number of distant background
galaxies\footnote{We have changed this limit up to
2600~km~s$^{-1}$ for the upcoming analysis of the virial zone
itself~\cite{kas2018:Kashibadze_n_en}.}.

Totally, 3995 objects from HyperLEDA and NED databases satisfy
these conditions. We have investigated visually all object images
in the Digital Sky Survey (DSS), Sloan Digital Sky Survey
(SDSS)~\cite{aba2009:Kashibadze_n_en} and Panoramic Survey
Telescope and Rapid Responce System
(PanSTARRS)~\cite{cha2016:Kashibadze_n_en} verifying and refining
galaxy morphological classification. As a result, we have excluded
1837 fake objects from this sample; all of them turned out to be
stars projected onto distant galaxies, different parts of one and
the same galaxy, or ambiguous optical identifications of radio
sources. It is worth accentuating that the noncritical, formal use
of basic data should lead to about 46\% of ``spam'' in the
reviewed sample, corrupting the further analysis.

We have obtained new distance estimates for 563 galaxies from this
list. 386 estimates have been resulted from the relation of Tully and
Pierce~\cite{tul2000:Kashibadze_n_en}:
$$M_B=7.27(2.5-\log W_{50})-19.99;$$
the hydrogen line widths are corrected for galaxy inclinations,
and the data on axial ratio have been extracted from HyperLEDA.
Distance estimates for the remaining 177 gas-rich dwarf galaxies
have been derived from the baryonic Tully--Fisher
relation~\cite{kar2017b:Kashibadze_n_en}. The authors have argued
that considering inclinations of dwarf galaxies does not
noticeably improve the dispersion on the Tully--Fisher diagram.
For that reason, we have not applied corrections for galaxy
inclinations in the case of baryonic TF-relation.

The resulting list of 2158 galaxies is presented in
Table~\ref{table01:Kashibadze_n_en} with the full machine readable
version available at SIMBAD Astronomical
Database\footnote{\protect\url{http://vizier.u-strasbg.fr/viz-bin/VizieR?-source=J/other/AstBu/73.124}}. The
table columns contain (1)~galaxy name; (2)~equatorial coordinates
J2000.0; (3)~supergalactic coordinates; (4)~radial velocity of a
galaxy (km~s$^{-1}$) relative to the Local Group centroid with
apex parameters adopted in NED; (5)~morphological type according
to de~Vaucouleurs classification; in addition to the
de~Vaucouleurs types we note also for compact objects: Ec (compact
ellipticals) and BCD (blue compact dwarfs); for dwarfs of low and
very low surface brightness we use the designations: Ed, Edn
(elliptical shape), Sph, Sphn (round); the symbol ``n'' means a
presence of a star-like nucleus; I, Ir (diffuse objects having no
blue knots); (6)~apparent $B$\mbox{-}magnitude from HyperLEDA or
NED, values without decimals indicate our visual estimates;
(7)~21-cm H\,I line width (km~s$^{-1}$) at half maximum;
(8)~apparent H\,I-magnitude \mbox {$m_{21} = 17.4-2.5\log F({\rm
H\,I})$}, where $F({\rm H\,I})$ is the H\,I line flow (Jy
km~s$^{-1}$); \mbox{(9), (10)}~distance modulus and galaxy
distance (Mpc); \linebreak (11)~method applied to determine
distance. We distinguished methods based on supernovae (SN) and
cepheids (cep) luminosity, the tip of the red giant branch (rgb)
recipe~\cite{lee1993:Kashibadze_n_en}, and the surface brightness
fluctuation (sbf) method~\cite{ton2001:Kashibadze_n_en}. The
distance estimates made by these techniques have an accuracy of
5\mbox{--}10\%, and we specify them as ``accurate'' hereafter. The
secondary or ``supplement'' distance estimates---from the
Tully--Fisher relation (TF or tf), the fundamental plane (FP) or
from globular cluster luminosity (gc), calibrated by cepheids,
have a typical accuracy of 20-25\%. We have marked Tully--Fisher
distances extracted from NED as ``tf'', while our new estimates
after  Tully and Pierce~\cite{tul2000:Kashibadze_n_en} are denoted
with capitals ``TF''; in the case of gas-rich dwarf galaxies their
distances after
Karachentsev~et~al.~\cite{kar2017b:Kashibadze_n_en} are labeled as
``TFb''. In total, galaxies with distance measurements amount to
1119, which is 52\% of the total sample. The catalog Cosmicflows-3
Distances~\cite{tul2016:Kashibadze_n_en} contains only 450
distance estimates in this region.

\begin{figure*} 
\includegraphics[width=\textwidth]{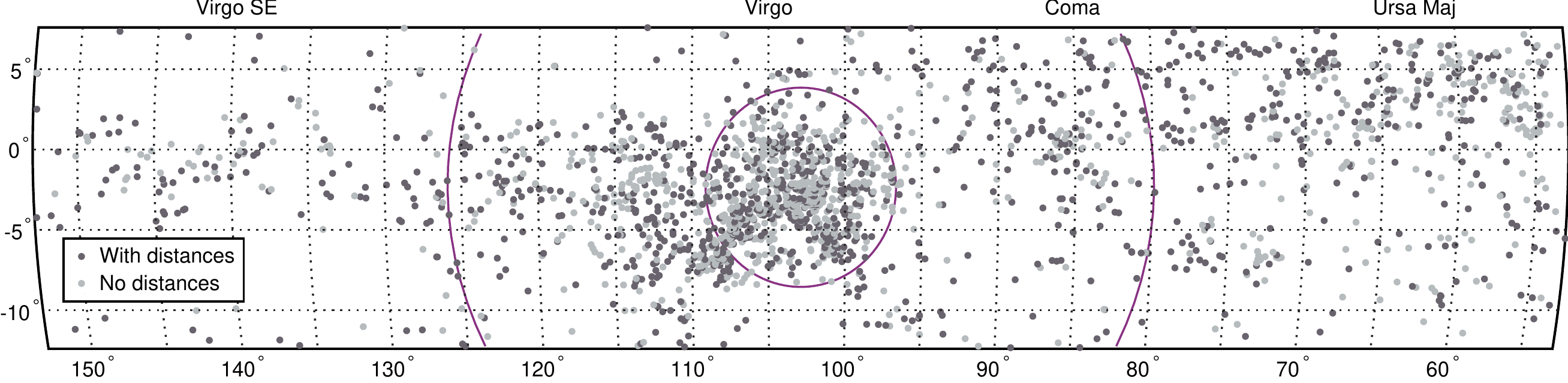}
\caption{Distribution of 2158 galaxies of the LSC belt with radial
velocities $V_{\rm LG}<2000$~km~s$^{-1}$ over the sky in
supergalactic coordinates. Galaxies with distance estimates are
marked by darker symbols. The central circle and the arcs denote the
virial radius (1.8~Mpc) and the radius of the zero velocity surface
(7.2~Mpc) for the Virgo cluster.} \label{fig02:Kashibadze_n_en}
\end{figure*}

The distribution of galaxies by their supergalactic coordinates is
presented in Fig.~\ref{fig02:Kashibadze_n_en}, where objects with measured distances
and without them are shown with dark and open markers, respectively.
The central circle with the radius of $6\fdg2$ outlines the virial
zone of the Virgo cluster (the virial radius $R_V=1.8$~Mpc), and the
arcs of the bigger circle $23\fdg6$ or $R_0=7.2$~Mpc separate the
infall region from the cosmological expansion. As it can be seen, the
distribution of galaxies outside the virial zone is notably
non-uniform and asymmetric, while the ratio of galaxies with measured
distances does not change significantly from the left (southern) side
of the belt to the right (northern) one. All three dynamical zones:
the virial core, the infall zone and the field are well represented
in our sample.

\begin{figure} 
\includegraphics[width=0.5\textwidth]{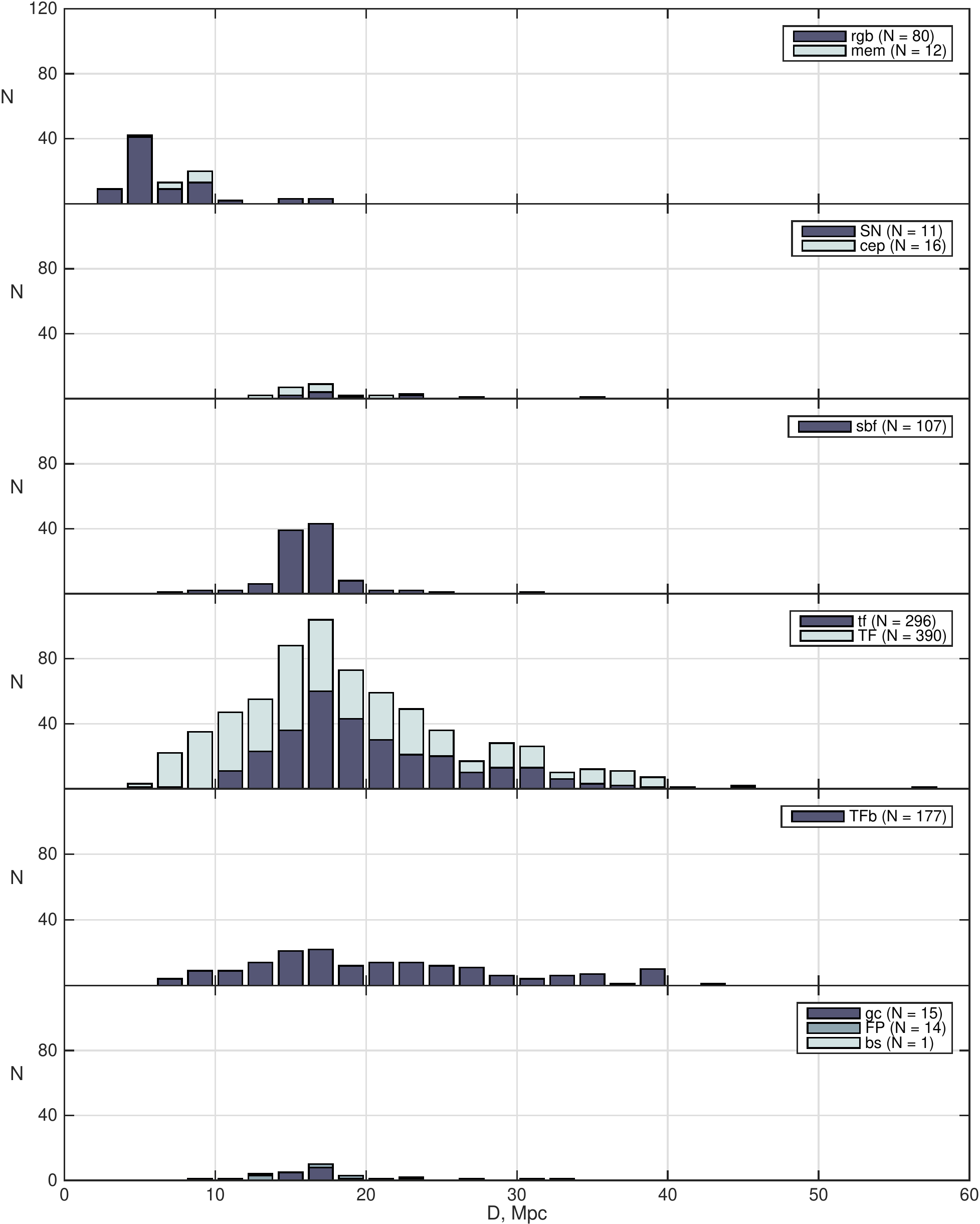}
\caption{Distribution of the LSC galaxies by distances measured from different
methods.}
\label{fig03:Kashibadze_n_en}
\end{figure}

Figure~\ref{fig03:Kashibadze_n_en} shows the distribution of LSC
belt ga\-la\-xies by their distances measured from different
me\-thods. Three top panels correspond to highly accurate methods.
80 galaxies with RGB distances are supplemented by 12 galaxies
with distances measured from their obvious membership (mem) in
groups with reliable distance estimates of other members,
ge\-ne\-ral\-ly more bright. Three bottom panels of
Fig.~\ref{fig03:Kashibadze_n_en} depict the distribution of
galaxies with less accurate distances. Comparison of these data
demonstrates the inherent difference between the applied methods
in their effective depth: galaxies with RGB distances are
concentrated mainly in the Local Volume ($D<11$ Mpc) while the
bulk of galaxies with Tully--Fisher distances resides behind the
Virgo cluster. The peak $N(D)$ values at $D\simeq16$~Mpc are
caused by the Virgo cluster members. However, gas-rich dwarfs with
``TFb'' distances do not show such a peak. These factors should be
considered when analyzing the infall of galaxies toward the Virgo
center from foreground and background of the cluster.

\begin{figure} 
\includegraphics[width=0.48\textwidth]{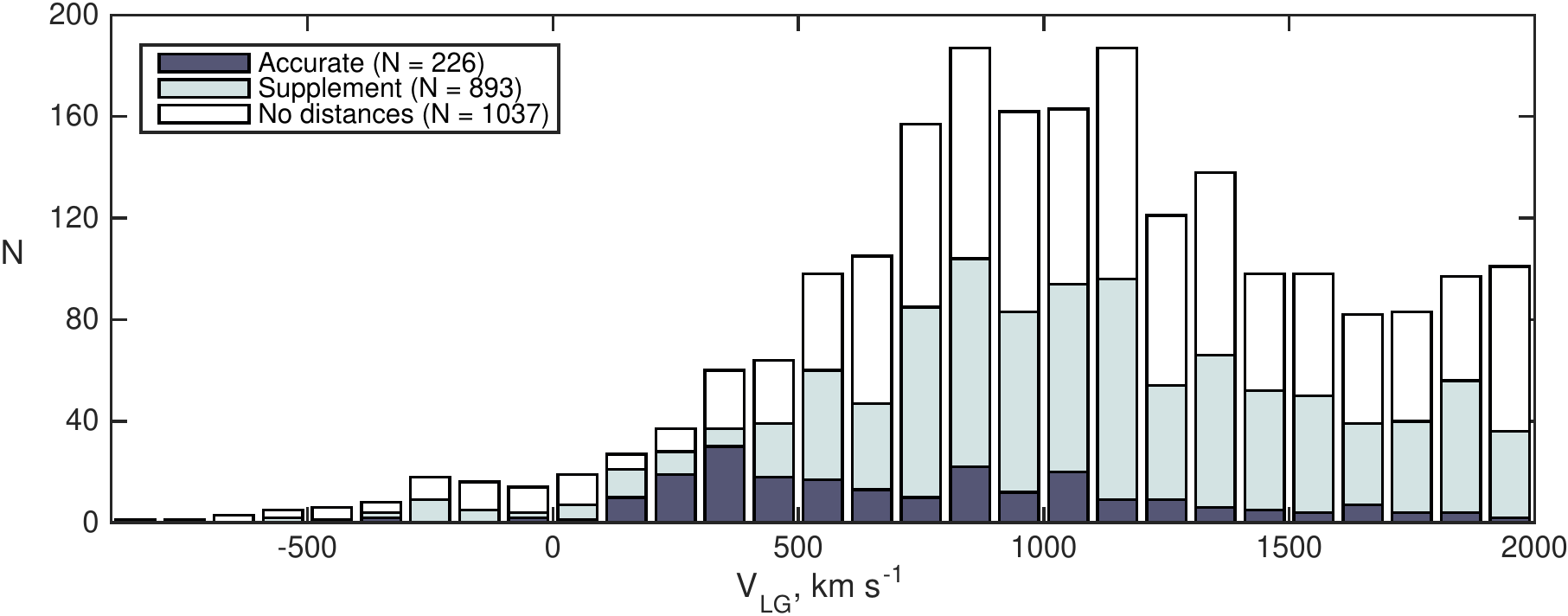}\\
\vspace{0.33cm}
\includegraphics[width=0.48\textwidth]{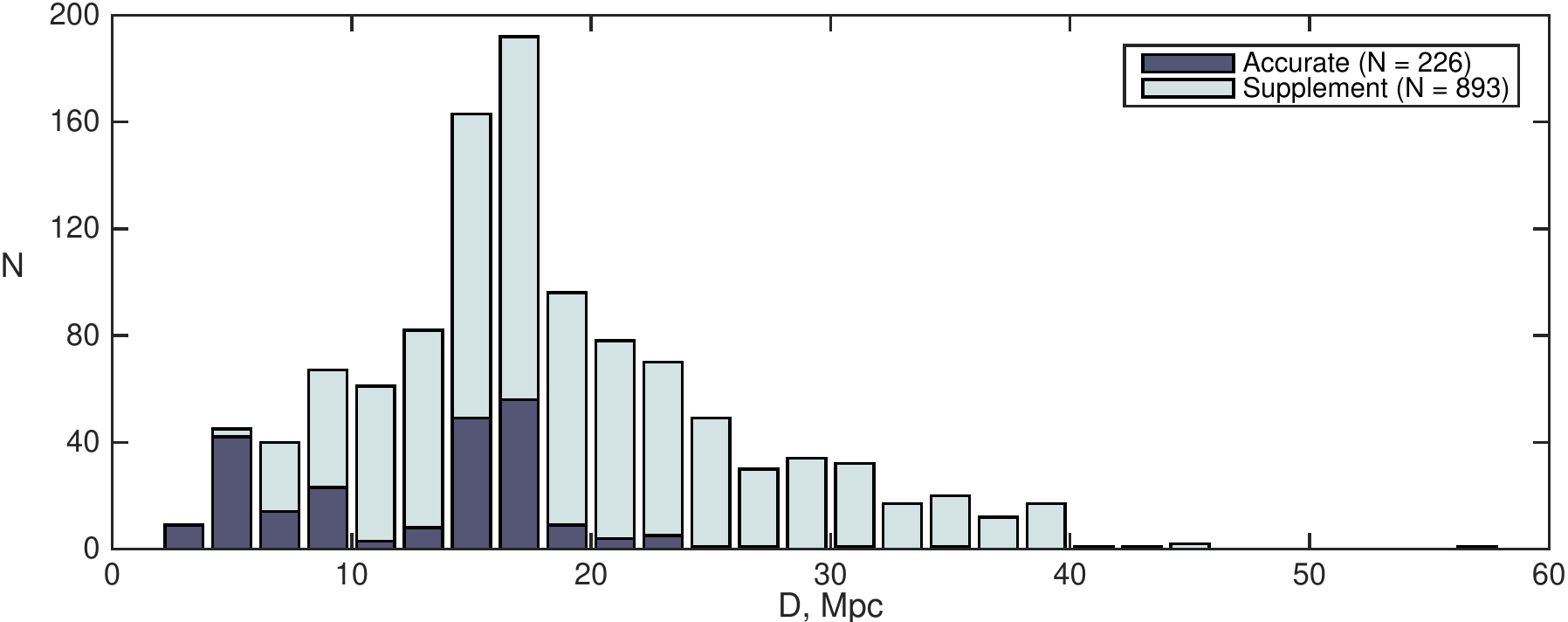}\\
\vspace{0.33cm}
\includegraphics[width=0.48\textwidth]{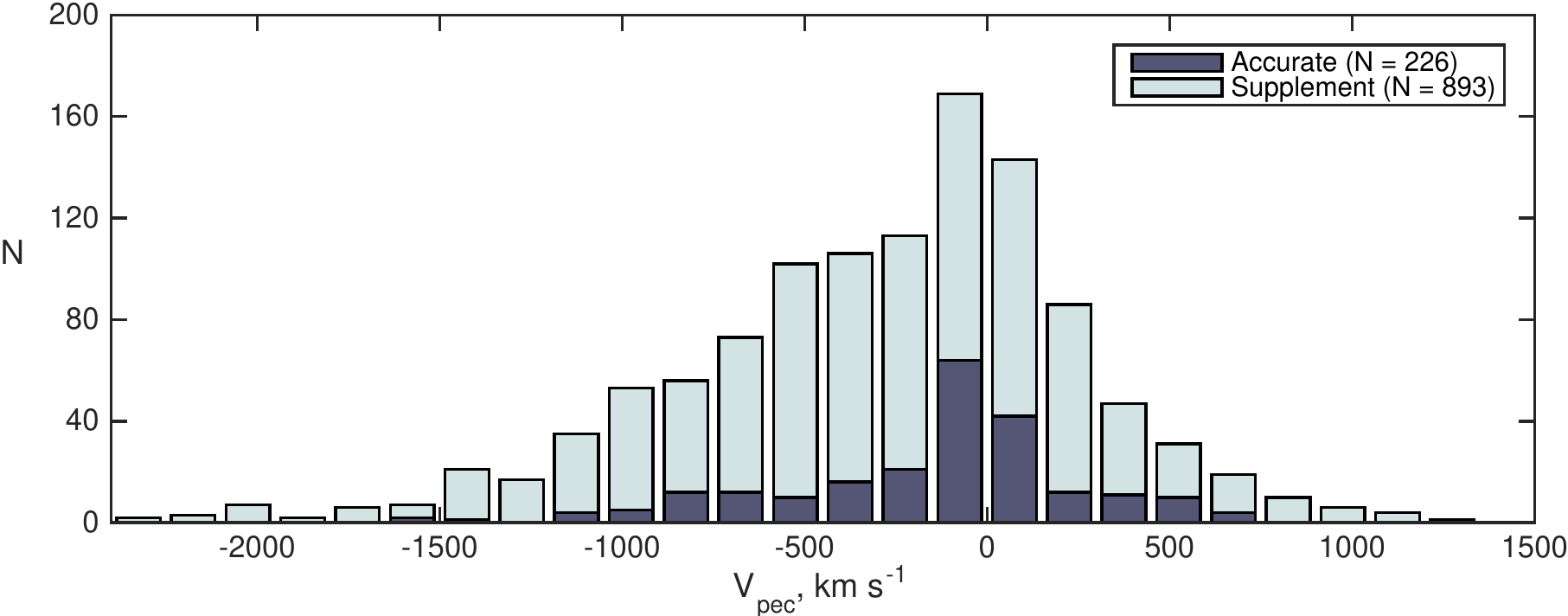}
\caption{Distribution of the LSC galaxies by (top) radial velocities, (middle)
distances and (bottom) peculiar velocities. Galaxies with accurate distances are
marked with dark grey.}
\label{fig04:Kashibadze_n_en}
\end{figure}

The general distribution of LSC belt galaxies by their radial
velocities, distances and peculiar velocities $V_{\rm pec}\!=\!
V_{\rm LG}\!-\!H_0 D$, where $H_0\!=\!73$~km~s$^{-1}$~Mpc$^{-1}$,
is presented in three panels of Fig.~\ref{fig04:Kashibadze_n_en}.
The wide peak $N(V_{\rm LG}$) with $V_{\rm
LG}=1000\pm500$~km~s$^{-1}$ is formed by members of the Virgo
cluster and the adjacent structures. The ratio of galaxies without
distance estimates grows faintly with radial velocities. As it can
be seen from the middle panel data, more than 99\% of galaxies
have distances within 40~Mpc. The tail area of distribution
($D=40$--60~Mpc) might be due to systematical errors in measured
widths $W_{50}$, like in the case of UGC\,6372.

Peculiar velocities of galaxies are distributed in the wide range of $[-2300,
+1300]$~km~s$^{-1}$ with the maximum near zero. The $N(V_{\rm pec})$ histogram looks
asymmetric and shifted to negative values $V_{\rm pec}$. The observed asymmetry
could have different reasons, such as:
 \begin{list}{}{
 \setlength\leftmargin{2mm} \setlength\topsep{2mm}
 \setlength\parsep{0mm} \setlength\itemsep{2mm} }
\item (a) the motion of the Local Group toward the Virgo center with
the velocity of  about 200~km~s$^{-1}$;%
\item (b) the limit of $V_{\rm LG}<2000$~km~s$^{-1}$ filtering out a
part of falling galaxies at the foreground side of the cluster but
passing ones that are moving from the background side;%
\item (c) the Malmquist bias manifesting itself when symmetrical
errors of distance moduli $\Delta(m-M)$ turn into asymmetrical errors
of distances $\Delta D$ (this effect is less noticeable for galaxies
with accurate distance estimates);%
\item (d) the galaxy associations with significant peculiar
velocities presenting in the LSC belt; in such a case the large
positive peculiar velocities are cut off by the condition $V_{\rm
LG}<2000$~km~s$^{-1}$ while the conspicuous negative peculiar
velocities satisfy this condition.
\end{list}

\section{OBSERVING TRENDS ALONG SGL}
\subsection{Kinematic/Geometric Trends}
In the standard cosmological model $\Lambda$CDM galaxy clusters are
formed as dense hub at the crossing of walls and filaments; along
these structures, galaxies fall into the virial zone of a cluster.
The distribution of galaxies by their distances and radial velocities
should have been impacted by galaxy flows along cosmic filaments.
Tully~et~al.~\cite{tul2016:Kashibadze_n_en} suggested that the main channel feeding
the virial zone of the Virgo cluster is the Virgo Southern Extension
filament (VirgoSE) at supergalactic latitudes ${\rm
SGL}>110^{\circ}$. Figures 5 and 6 show the distribution of
velocities and distances of galaxies along SGL. Taking into account
the selection effects listed above, we have divided the LSC belt
galaxies into two subsamples having accurate and secondary distance
indicators. The bold line reflects the running median with smoothing
window of $\Delta {\rm SGL}=4^{\circ}$, and narrow lines above and
below correspond to running quartiles.

\begin{figure} 
\includegraphics[width=0.48\textwidth]{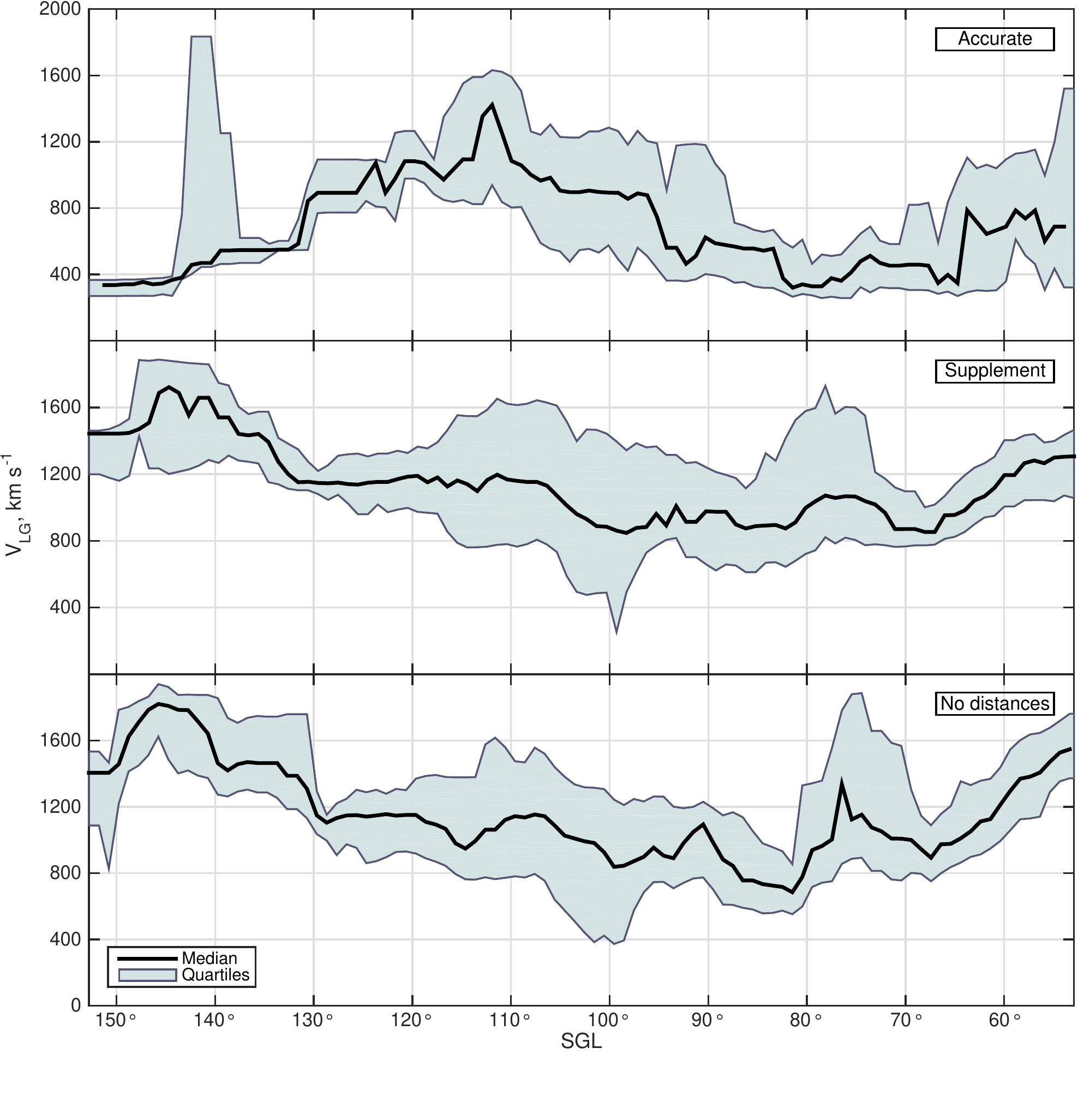}
\caption{Running median and quartiles of radial velocities along the LSC belt
for galaxies with measured (top, middle) and unknown (bottom) distances.}
\label{fig05:Kashibadze_n_en}
\end{figure}

\begin{figure}[tp] 
\includegraphics[width=0.48\textwidth]{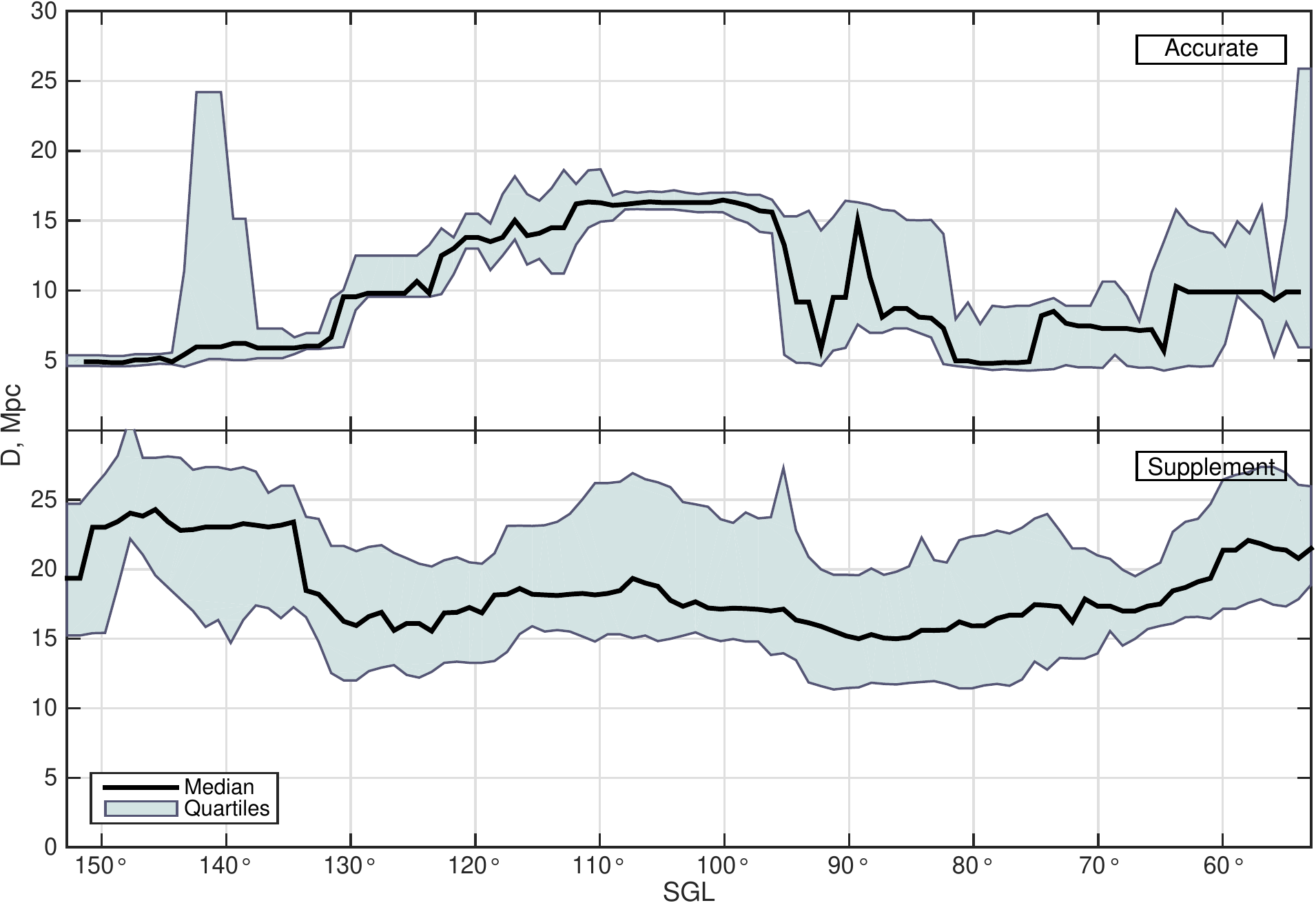}
\caption{Running median and quartiles of distances along the supergalactic
longitude.}
\label{fig06:Kashibadze_n_en}
\end{figure}

As it can be seen, in the case of ``accurate'' distances (top
panels of Figs.~\ref{fig05:Kashibadze_n_en} and
\ref{fig06:Kashibadze_n_en}) the median velocity and the median
distance tend to increase from the belt ends to its center fixed
by the Virgo cluster. This is just what should be expected if
galaxies move along the southern and northern filaments adjoining
the Virgo cluster by their far ends. However, this phenomenon
could also be caused by the selection effect: Ferrarese et
al.~\cite{fer2012:Kashibadze_n_en} and
Mei~et~al.~\cite{mei2007:Kashibadze_n_en} carried out special
efforts to measure distances of galaxies in the Virgo core by RGB
and SBF methods, while at the ends of the belt RGB distances are
available only for members of the nearest groups (NGC\,4244,
NGC\,4631, NGC\,5236). Another pattern is demonstrated by the
running median in the case of TF distances (middle panel of
Fig.~\ref{fig05:Kashibadze_n_en}). Note also that the drift of
$V_{\rm LG}$ running median for galaxies without distance
estimates (bottom panel of Fig.~\ref{fig05:Kashibadze_n_en})
repeats roughly the shape of the median in the middle panel.

As it follows from the data presented at the top panel of
Fig.~\ref{fig06:Kashibadze_n_en}, the scatter of accurate
distances in the virial zone of the cluster (\mbox{${\rm
SGL}=97-110^{\circ}$}) is remarkably small and comparable with the
virial radius of the cluster. The band-average interquartile width
for accurate distances is 2 or 3 times less than the similar width
for supplement distances, cor\-res\-pon\-ding roughly to the
accuracy ratio for these methods (5--10\% vs. \mbox{20--25\%}).

\begin{figure}[bp] 
\includegraphics[width=0.48\textwidth]{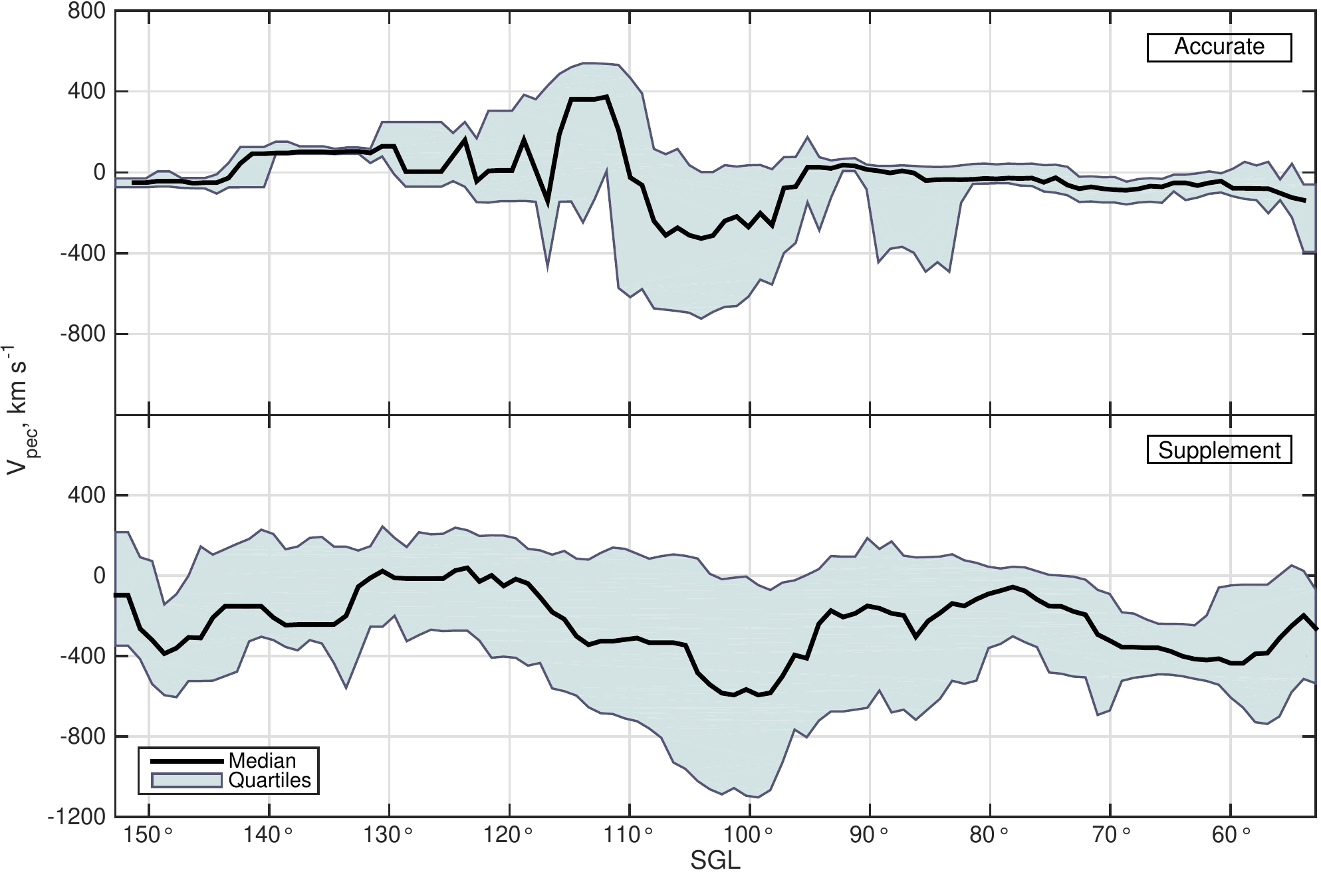}
\caption{Running median and quartiles of peculiar velocities along the LSC belt.}
\label{fig07:Kashibadze_n_en}
\end{figure}

Figure~\ref{fig07:Kashibadze_n_en} shows the running median and
quartiles for peculiar velocities of galaxies along supergalactic
longitude. At the ends of the belt ${\rm SGL}<80^{\circ}$ and
${\rm SGL}>130^{\circ}$ galaxies with accurate distances
demonstrate insignificant shift and small scatter of peculiar
velocities. This is caused by the fact that most of them belongs
to the ``cold'' Local Sheet where the random non-Hubble velocities
are minor. On the contrary, galaxies with TF distances are
situated mainly at the far end of the volume external to the Local
Sheet, having predominantly negative peculiar velocities. Both
subsamples in the top and bottom panels detect the predicted
growth of $V_{\rm pec}$ dispersion in the center of the belt due
to the virial motions in the cluster.

Another feature being seen in the top panel of
Fig.~\ref{fig07:Kashibadze_n_en} is the excess of galaxies with
positive peculiar velocities near the southern side of the Virgo
cluster $({\rm SGL}\simeq113^{\circ})$. It is probably caused by
the association of galaxies around NGC\,4527 situated at the
proximal side of the Virgo cluster and falling toward its center.
Yet another feature---the zone of negative values of $V_{\rm
pec}$---can be seen in the interval of \mbox{${\rm
SGL}=83^{\circ}$--$90^{\circ}$}. It corresponds to the velocity
anomaly in Coma\,I~\cite{kar2011:Kashibadze_n_en} and will be
discussed in more detail below.

As it follows from the data presented at both panels of
Fig.~\ref{fig07:Kashibadze_n_en}, the general trend of the median peculiar velocity
along the belt is imperceptible as expected in the case when
filamentary structures near the Virgo cluster are oriented almost
perpendicular to the line of sight.

\subsection{Morphological Landscape}
\begin{figure*} 
\includegraphics[width=0.9\textwidth]{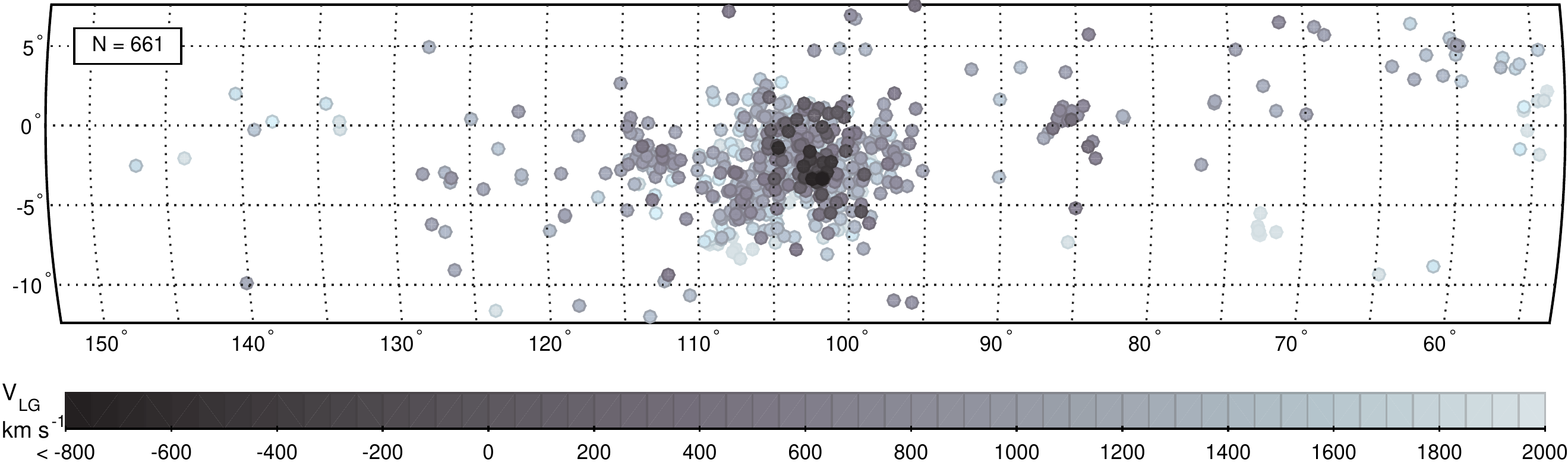}
\vspace{0.5cm}
\includegraphics[width=0.9\textwidth]{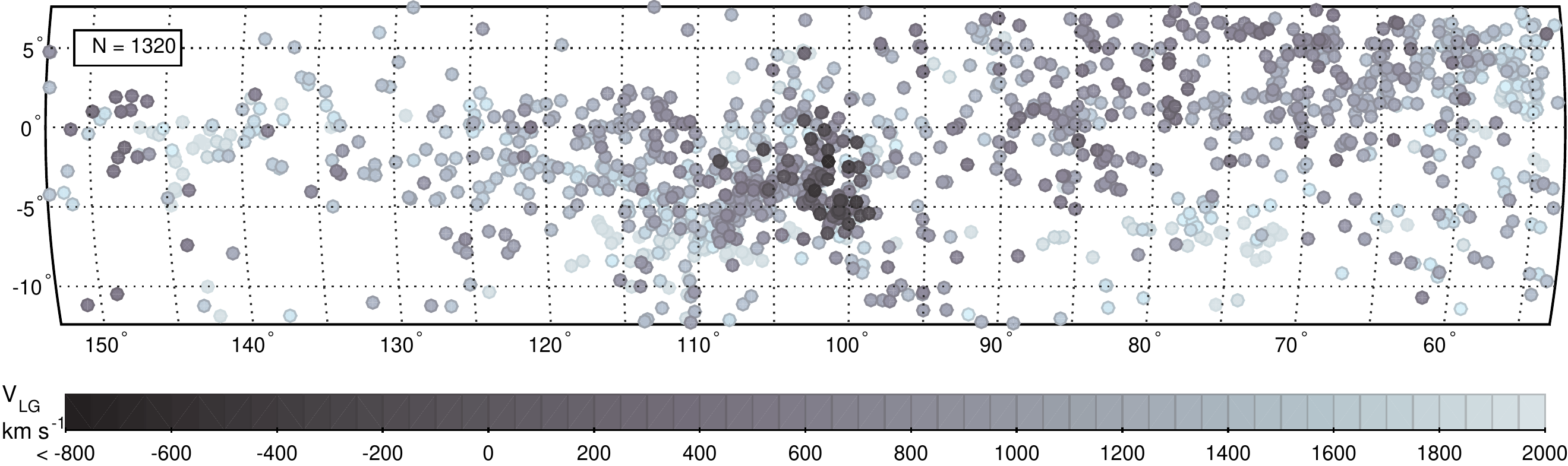}
\vspace{0.5cm}
\includegraphics[width=0.9\textwidth]{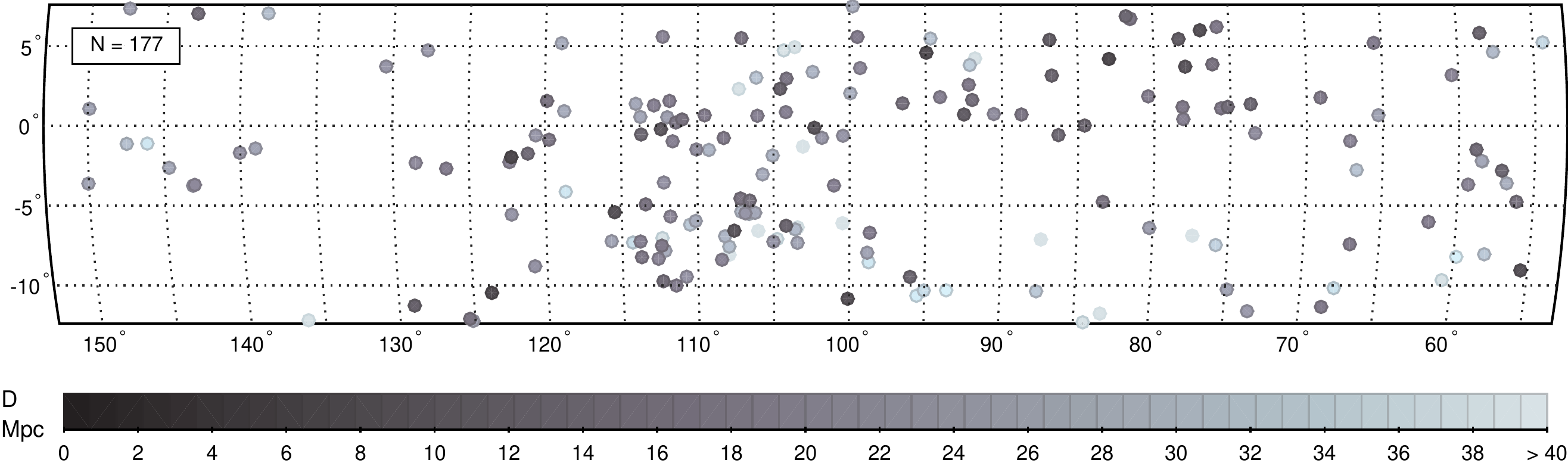}
\caption{Distribution of (top) early type galaxies, (middle) late type galaxies,
(bottom) gas-rich dwarfs along the LSC belt. Radial velocity or distance scales
are given below each panel.}
\label{fig08:Kashibadze_n_en}
\end{figure*}

The distribution of LSC belt galaxies by their morphological types
is shown in Fig.~\ref{fig08:Kashibadze_n_en}; the radial velocity
or distance scales are given below each panel. The early type
galaxies demonstrate strong concentration toward the core of the
Virgo cluster (top panel of Fig.~\ref{fig08:Kashibadze_n_en}).
About 3/4 of them are located inside the virial radius of the
cluster, and the others are associated with galaxy groups. The
equatorial group with ${\rm SGL}\simeq86^{\circ}$ corresponds to
the Coma\,I cloud around NGC\,4278 with significant negative
peculiar velocities.

The distribution of the late type galaxies in the middle panel shows much lesser
concentration toward the virial zone of the Virgo cluster. This effect is well
known as morphological segregation reflecting the density of the environment.

Haynes and Giovanelli~\cite{hay1984:Kashibadze_n_en} have pointed
out that spiral galaxies of a certain morphological type located
in the Virgo cluster core are characterized with H\,I deficiency
relative to the field galaxies of the same type. Ram pressure
stripping of gas in the dense virial zone of the cluster is the
evident scheme explaining the H\,I deficiency.\footnote{Planck
Collaboration~\cite{ade2016:Kashibadze_n_en} have reported on
detection of the Sunyaev--Zeldovich effect from the Virgo cluster.
They estimate the mass of the hot intracluster gas as
\mbox{$(1.5\pm0.1)\times 10^{14}M_{\odot}$}, which is 17 (!) times
more than the total stellar mass of the cluster.} The stripping
process is most effective for irregular dwarf galaxies because of
their shallow potential wells. The bottom panel of
Fig.~\ref{fig08:Kashibadze_n_en} represents the distribution of
the gas-rich dwarfs ($m_{21}<B_T$) with baryonic Tully--Fisher
distances. The concentration of these objects toward the Virgo
cluster is hardly detectable providing a detached perspective of
the H\,I deficiency phenomenon. Some lack of uniformity in the
distribution of gas-rich galaxies over the considered region is
caused by the borders of the HIPASS and ALFALFA H\,I surveys.

\subsection{Nearby Groups in the LSC Band}
Makarov and Karachentsev~\cite{mak2011:Kashibadze_n_en} (MK\,11)
have published a list of 350 groups consisting of galaxies with
radial velocities $V_{\rm LG}<3500$~km~s$^{-1}$. This all-sky
sample includes 50 groups with coordinates falling into the LSC
belt and having mean radial velocity $V_{\rm
LG}<2000$~km~s$^{-1}$. These groups are listed in
Table~\ref{table02:Kashibadze_n_en}. Its columns contains (1)~name
of the brightest member; (2), (3)~supergalactic coordinates of the
group center; (4)~number of galaxies with measured radial
velocities; (5)~mean radial velocity of a group (km~s$^{-1}$);
(6)~mean group distance (Mpc); (7)~peculiar velocity of a group
(km~s$^{-1}$).

It should be noted that MK\,11 galaxy clustering criterion used
velocity based distance estimates, as there were only few direct
distance measurements at that time. This circumstance has caused
some uncertainty in clustering results for the Virgo virial zone
characterized by large motions. The bulk of Virgo galaxies has
been clustered around giant elliptical NGC\,4472 instead of
NGC\,4486. Moreover, several galaxy groups with small radial
velocities (NGC\,4216, NGC\,4342, NGC\,4402, NGC\,4552) turned out
to be fake structures.

\begin{figure*} 
\includegraphics[width=0.9\textwidth]{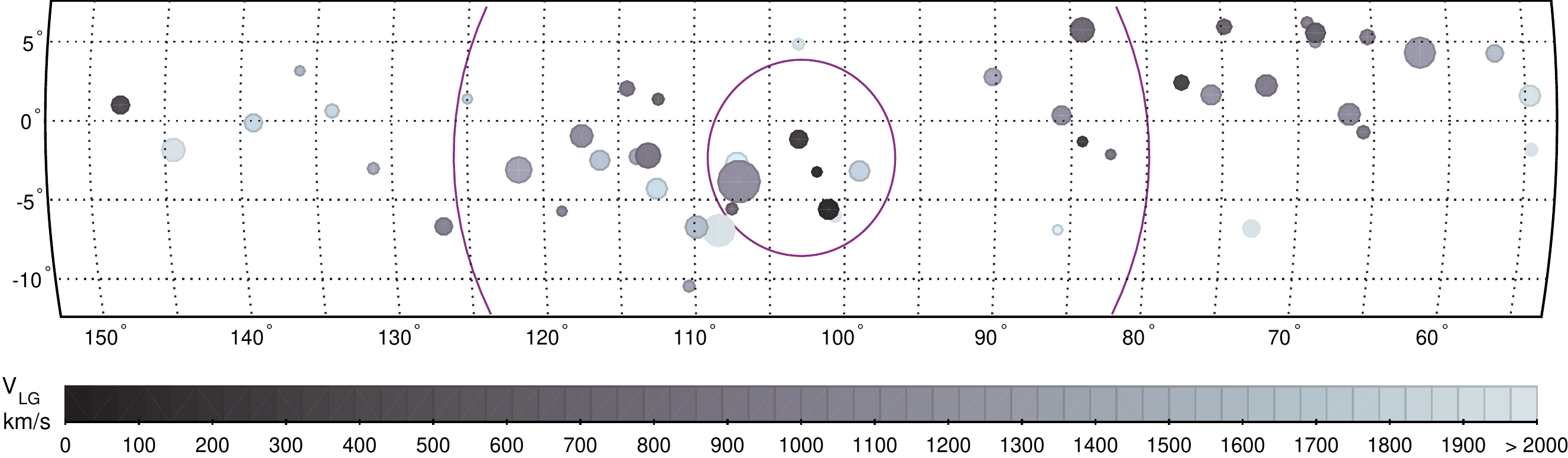}
\caption{Galaxy groups in the considered region. The marker size is proportional
to logarithmic number of galaxies with measured radial velocities; the density
specifies mean radial velocity of a group according to the scale given below.
The large and small circles correspond to the radius of the zero velocity
surface and the virial radius of the Virgo cluster.}
\label{fig09:Kashibadze_n_en}
\end{figure*}

Figure~\ref{fig09:Kashibadze_n_en} shows the distribution of 50
MK\,11 groups marked as circles over the considered region. The
size of a circle is proportional to logarithmic number of galaxies
with measured radial velocities, and the color reflects mean
radial velocity of a group according to the scale given below.

As it follows from the data presented in the last column of
Table~\ref{table02:Kashibadze_n_en}, some galaxy groups have
significant peculiar velocities $|V_{\rm pec}|>500$~km~s$^{-1}$:
$-$749~km~s$^{-1}$ (NGC\,3838), $-$518~km~s$^{-1}$ (NGC\,3900),
$-$972~km~s$^{-1}$ (NGC\,4150), +534~km~s$^{-1}$ (NGC\,4527),
$-$542~km~s$^{-1}$ (NGC\,4636), $-$754~km~s$^{-1}$ (NGC\,4900). If
the individual distance error is equal to ap\-pro\-xi\-ma\-tely
20\% and the group includes more than four members with distance
estimates, then the expected peculiar velocity error makes less
than 150~km~s$^{-1}$ for the typical distance of nearly 20~Mpc.
Hence, the conspicuous peculiar velocities of several groups in
the vicinity of Virgo have physical origins and do not result from
errors in distance measurements.

\section{VELOCITY ANOMALY IN COMA\,I}
The most spectacular example of significant peculiar velocities is
the galaxy association near NGC\,4278 known as Coma\,I cloud.
Considering probable Local Volume galaxies with velocities $V_{\rm
LG}<600$~km~s$^{-1}$,
Karachentsev~et~al.~\cite{kar2011:Kashibadze_n_en} have noticed
that the most distant ones ($D>13$~Mpc) cover the sky in quite a
spotty manner. New data originating from ALFALFA H\,I survey make
this pattern even more contrast.

\begin{figure*} 
\includegraphics[width=0.9\textwidth]{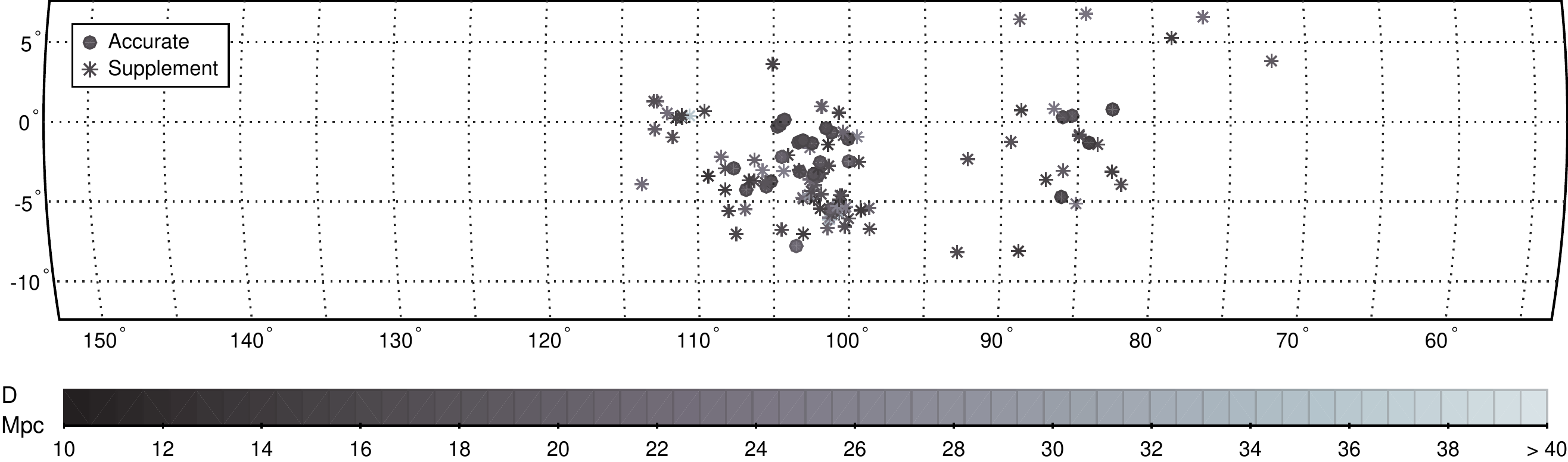}\\
\vspace{0.85cm}
\includegraphics[width=0.45\textwidth]{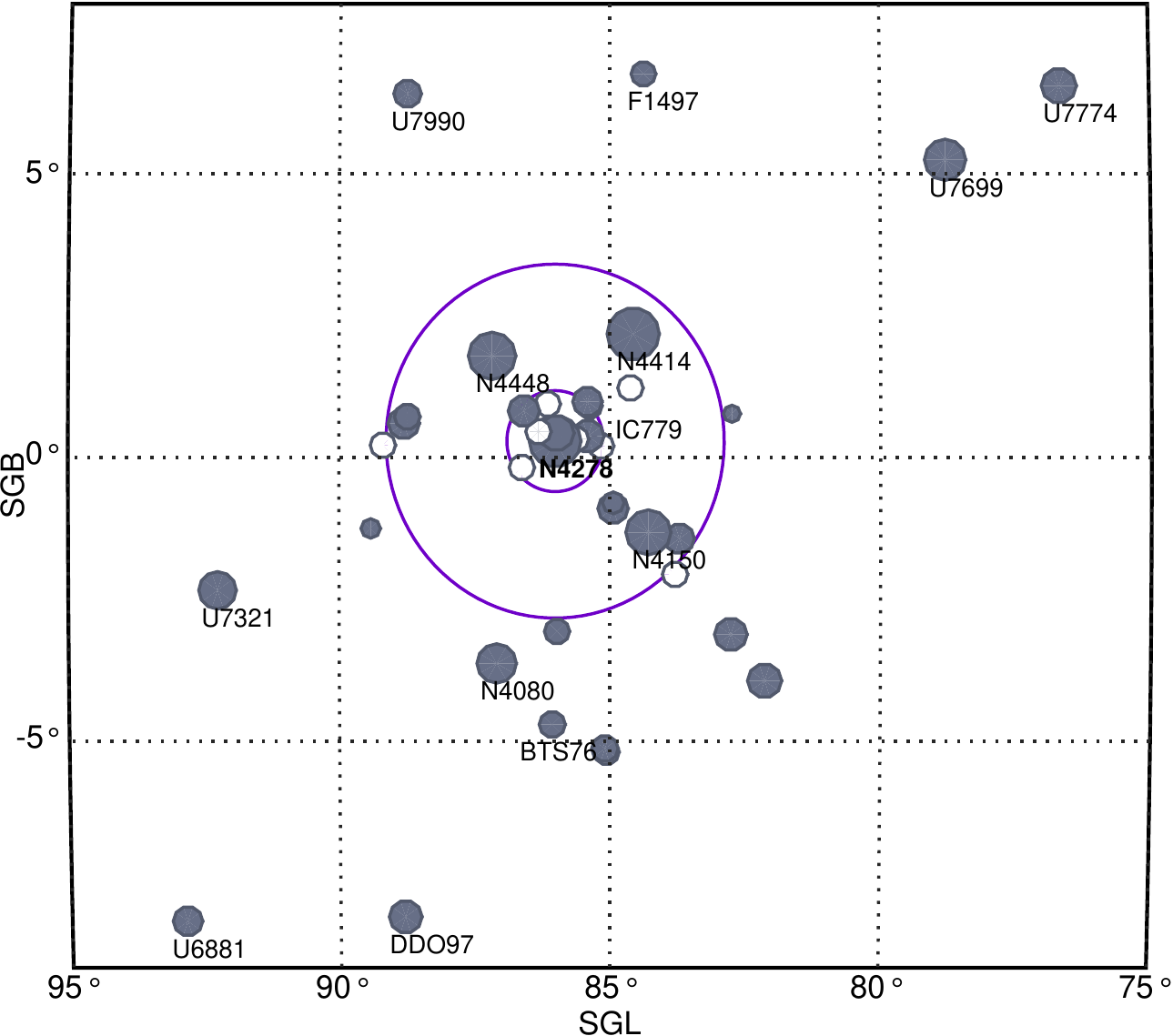}
\caption{Distribution of LSC belt galaxies with distances
$D>13$~Mpc and radial velocities $V_{\rm LG}<600$~km~s$^{-1}$ (top
panel). Galaxies with ``accurate'' and ``supplement'' distances
are marked by circles and asterisks, respectively. Bottom panel:
distribution of Coma\,I cloud galaxies around NGC\,4278 in
supergalactic coordinates. The marker size is proportional to
logarithmic stellar mass of a galaxy. Open circles denote galaxies
without individual distance estimates.}
\label{fig10:Kashibadze_n_en}
\end{figure*}

The distribution of galaxies with \mbox{$V_{\rm
LG}\!<\!600$~km\,s$^{-1}$} and $D>13$~Mpc over the LSC belt is
presented in the top panel of Fig.~\ref{fig10:Kashibadze_n_en}.
Galaxies with accurate and less precise distances are depicted,
respectively, as circles and asterisks. Predictably, the center of
this image is filled with Virgo members having large virial
velocities directed toward the observer. Besides the Virgo
cluster, the diagram displays the only structure ap\-pea\-ring as
a galaxy cloud around the massive elliptical galaxy NGC\,4278 near
${\rm SGL}=86^{\circ}$ and \mbox {${\rm SGB}=0^{\circ}$}. The
probability of all 23 galaxies outside the Virgo core laying
rightward does not exceed $10^{-6}$.

We have considered the wide outskirts of \linebreak NGC\,4278 and
distinguished 28 distant galaxies with small radial velocities as
possible members of the Coma\,I cloud.
We have enlarged this sample with other ten galaxies which have
not yet distance estimates but can still belong to this cloud, by
reference to their texture. The whole sample is presented in the
Table~\ref{table03:Kashibadze_n_en}. Its columns contain
(1)~galaxy name; (2)~equatorial coordinates; (3)~de~Vaucouleurs
numerical morphological type; (4)~radial velocity (km~s$^{-1}$);
(5), (6)~galaxy distance (Mpc) and the method applied to determine
it; (7)~apparent \linebreak \mbox {$B$-magnitude}; (8)~peculiar
velocity (km~s$^{-1}$). The distribution of these galaxies is
shown in the bottom panel of Fig.~\ref{fig10:Kashibadze_n_en}.
Galaxies with measured distances are marked solid while galaxies
without distance estimates are represented as open markers. Small
and large circles specify the virial radius (270~kpc) and the
radius of the zero-velocity surface (1.0~Mpc) around the NGC\,4278
taking its distance to be 16.1~Mpc.

As one can see, the Coma\,I cloud is a rather loose formation with
only 1/3 of the galaxies inside the virial radius. On the other
hand, the presence of early type galaxies $(T<0)$ in the Coma\,I
cloud indicate the later stage of the group evolution.
Pro\-ba\-bly, NGC\,4278 with its satellites forms a part of an
elongated fi\-la\-ment or sheet with the mean peculiar velocity of
$-$840~km~s$^{-1}$, which is comparable with virial velocities in
rich galaxy clusters.

The Coma\,I cloud is situated at the angular distance of about
$18^\circ$ from the Virgo center, i.e. near the radius of its
zero-velocity surface ($23^\circ$). The distances of both
structures are roughly equal: 16.1~Mpc and 16.7~Mpc. Thus, the
expected radial infall of the Coma\,I cloud to the Virgo cluster
supposes its motion to be close to tangential. In this case the
Coma\,I peculiar velocity should be insignificant.

Earlier we have made a suggestion~\cite{kar2011:Kashibadze_n_en}
that conspicuous peculiar motions in the Coma\,I region could be
associated with a dark attractor having mass of about $2\times
10^{14}M_{\odot}$. Another explanation of this puzzling phenomenon
assumes a fast coherent motion of the whole filament with the
NGC\,4278 group toward the observer. But the source of such a high
velocity stays unclear. Note, however, that the mean peculiar
velocity of the Coma\,I cloud in the cosmic microwave background
(CMB) reference frame appears to be appreciably lower
($-$526~km~s$^{-1}$) as compared to its peculiar velocity in the
Local Group reference frame ($-$840~km~s$^{-1}$).

\section{VIRGOCENTRIC INFALL}
The neighboring massive Virgo cluster produces significant
distortions in the velocity field of nearby galaxies relative to
the undisturbed Hubble flow. The dynamical models and the
observational evidences of this effect known as Virgocentric
infall were studied by
Hoffman~et~al.~\cite{hof1980:Kashibadze_n_en}, Tonry and
Davis~\cite{ton1981:Kashibadze_n_en}, Hoffman and
Salpeter~\cite{hof1982:Kashibadze_n_en}, Tully and
Shaya~\cite{tul1984:Kashibadze_n_en},
Teerikorpi~et~al.~\cite{tee1992:Kashibadze_n_en},
Ekholm~et~al.~\cite{ekh1999:Kashibadze_n_en},
Tonry~et~al.~\cite{ton2000:Kashibadze_n_en,ton2001:Kashibadze_n_en},
and Karachentsev and Nasonova~\cite{kar2010:Kashibadze_n_en}.
Using the highly accurate Hubble Space Telescope distance
measurements for galaxies located near the proximal side of the
Virgo cluster, Karachentsev~et~al.~\cite{kar2014b:Kashibadze_n_en}
have determined the radius of the zero-velocity surface separating
the infall region from the global cosmic expansion to be
$R_0=7.2\pm0.7$~Mpc. The total mass of the cluster inside the
$R_0$ radius is expressed as
$$M_T=(\pi^2/8G) R^3_0 H^2_0/f^2(\Omega_m),$$
where the dimensionless parameter
\begin{eqnarray*}
f(\Omega_m)&=&(1-\Omega_m)^{-1}\\[-5pt]
&-&\dfrac{\Omega_m}{2}
(1-\Omega_m)^{-\dfrac{3}{2}} \cosh^{-1}(\dfrac{2}{\Omega_m}-1)
\end{eqnarray*}
changes in the range from 1 to 2/3 while varying $\Omega_m$ from 0
to 1. In the standard cosmological model with $\Omega_m=0.24$ and
$H_0=73$~km~s$^{-1}$~Mpc$^{-1}$, it corresponds to the total mass
of the cluster\linebreak \mbox{$M_T=(8.3\pm2.3)\times
10^{14}M_{\odot}$} matching closely the virial mass estimates:
$6.2\times 10^{14}M_{\odot}$~\cite{vau1960:Kashibadze_n_en},
\linebreak\mbox{$7.5\times
10^{14}M_{\odot}$}~\cite{tul1984:Kashibadze_n_en}
and $7.2\times 10^{14}M_{\odot}$~\cite{gir1999:Kashibadze_n_en}. 

\begin{figure*} 
\includegraphics[width=0.9\textwidth]{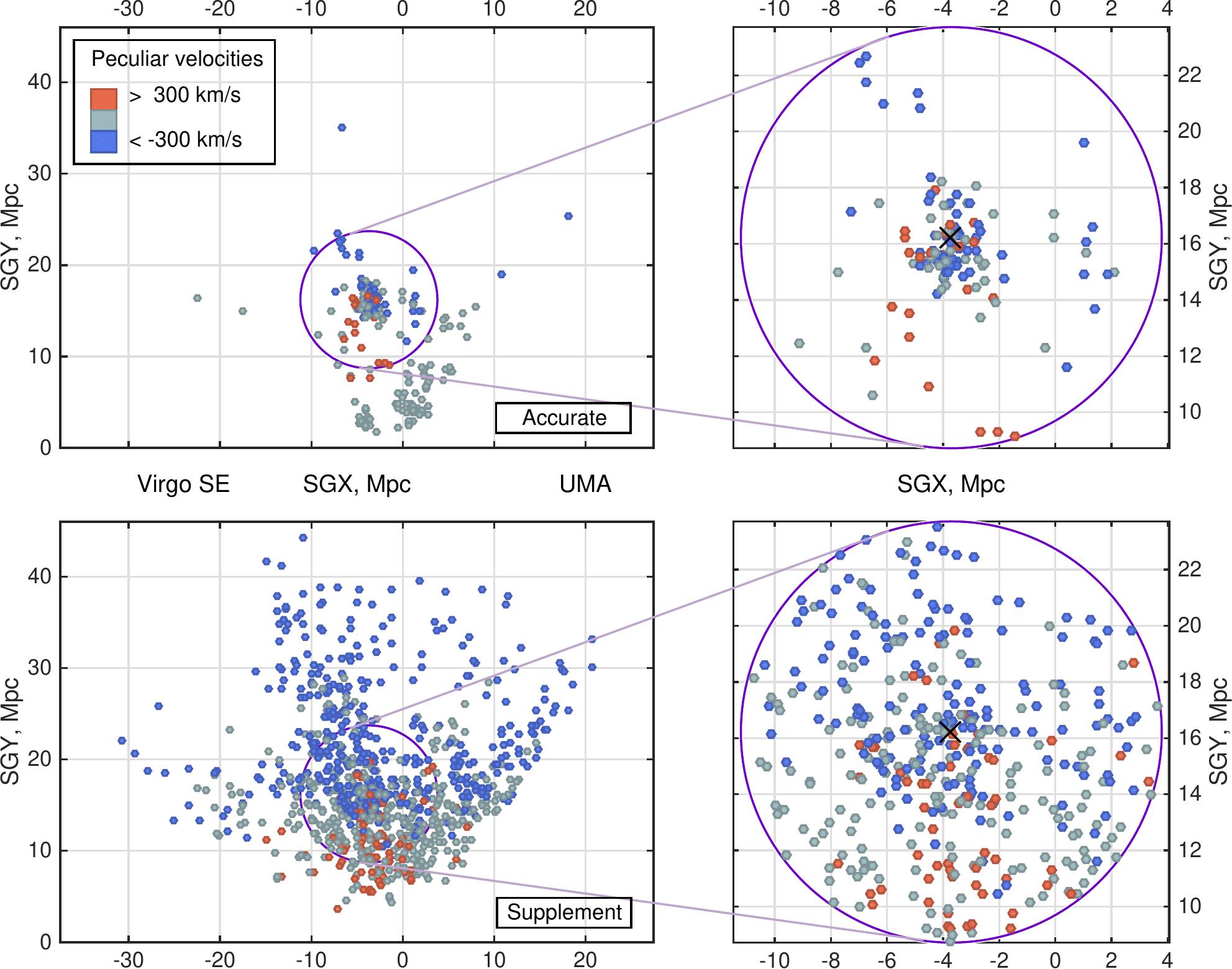}
\caption{Distribution of galaxies with large and small peculiar velocities in
Cartesian supergalactic coordinates. The circle corresponds to the radius of the
zero velocity surface.}
\label{fig11:Kashibadze_n_en}
\end{figure*}

The new data on distances of galaxies situated along the Local
Supercluster equator give the reason to reestimate $R_0$ and $M_T$
parameters for the Virgo cluster.
Figure~\ref{fig11:Kashibadze_n_en} presents the distribution of
the LSC belt galaxies projected onto the Local Supercluster plane
in Cartesian supergalactic coordinates. The top and the bottom
panels correspond to the ``accurate'' and ``supplement''
subsamples. The galaxy landscape inside the $R_0$ radius is inset
rightward. Red and blue markers indicate galaxies with peculiar
velocities $V_{\rm pec}<-300$~km~s$^{-1}$ and \mbox{$V_{\rm
pec}>+300$~km~s$^{-1}$}, respectively. Galaxies with insignificant
(intermediate) peculiar velocities are marked with grey.

The spherical infall pattern supposes nearby ga\-la\-xies with
positive peculiar velocities to be situated at the foreground side
of the Virgo cluster while ga\-la\-xies having negative peculiar
velocities should lie at its distant side. Such an effect is
actually observed in Fig.~\ref{fig11:Kashibadze_n_en} for both
subsamples. Several galaxies in the bottom panel deviate from this
regularity since the typical error for their Tully--Fisher
distances is 3\mbox{--}4~Mpc, being comparable with the virial
diameter of the Virgo cluster (3.6~Mpc).

``Grey'' galaxies with $|V_{\rm pec}|<300$~km~s$^{-1}$ in the top
panel form the Local Sheet elongated in the direction of Ursa
Major. The size of this ``cold'' domain amounts to about
$5\times15$~Mpc. The bottom, more populated panel manifests some
signs of an empty sector behind the Virgo cluster spreading toward
the Virgo Southern Extension. The data presented in the Table~\ref{table01:Kashibadze_n_en}
are obviously insufficient to detect the distant side of this
void.

The overrepresentation of galaxies with negative peculiar velocities
in the Fig.~\ref{fig11:Kashibadze_n_en} is caused both by infall of the Local Sheet
toward the Virgo cluster with the velocity of approximately
$200$~km~s$^{-1}$ and by bulk motion of galaxies behind the cluster
toward its center, i.e. in the direction of the observer.

To estimate the radius $R_0$ of the zero velocity surface, we have calculated
virgocentric distances of galaxies
$$D_{\rm vc}^2=D_g^2+D_c^2-2D_g D_c \cos \Theta, \eqno(1) $$
where $D_g$ and $D_c$ are the distances from the observer to a
galaxy and to the Virgo center respectively, and $\Theta$ is the
angular distance of a galaxy from the Virgo center. The velocity
of a galaxy relative to the cluster center has been defined within
the ``minor attractor'' model:
$$ V_{\rm vc}=V_g \cos\lambda -V_c\cos(\lambda+\Theta), \eqno(2) $$
where $\lambda$ is the angle between the line of sight and the
direction from the cluster center to a galaxy: \mbox
{$\tan\lambda=D_c\sin\Theta/(D_g-D_c\cos\Theta)$}, or within the
``major attractor'' model
$$V_{\rm vc}=[V_c \cos\Theta -V_g]/\cos\lambda. \eqno(3)$$

The first case supposes that the peculiar velocities of galaxies
are small compared to the regular Hubble flow. The second case
implies that the infall velocity predominates for most galaxies
(see details in~\cite{kar2010:Kashibadze_n_en}). The difference
between two models becomes in\-si\-gni\-fi\-cant if a galaxy is
located strictly behind $\lambda\simeq0$ or in front of
$\lambda\simeq180^\circ$ the cluster center.

For relating $V_{\rm vc}$ to $D_{\rm vc}$ it is essential to fix
both the distance and the radial velocity of the Virgo center.
According to Mei~et~al.~\cite{mei2007:Kashibadze_n_en},
$D_c=16.7\pm0.2$~Mpc.

Fixing $V_c$ is a more complicated problem. The mean radial velocity
and the radial velocity dispersion are known to depend on the
morphological composition of the sample~\cite{bin1993:Kashibadze_n_en}. Since
different methods of distance measurements use morphologically
diverse samples, and because of the morphological segregation in the
cluster, various $V_c$ estimation approaches suffer severely from
inconsistency. Additional complications appear due to the presence of
substructures forming the Virgo cluster: M\,87 cluster, M\,49
cluster, being projected onto the more distant structures:  W~cloud,
M~cloud, NGC\,4636 group~\cite{kou2017:Kashibadze_n_en,kim2014:Kashibadze_n_en}.

A modest summary of $V_c$ estimates known from literature is
presented in the upper part of the
Table~\ref{table04:Kashibadze_n_en}. Some its values are estimated
only for the Virgo core while others correspond to the whole
cluster. The number of galaxies in individual samples differ by an
order. Binggeli~et~al.~\cite{bin1993:Kashibadze_n_en} have used
only radial velocity data, while
Mei~et~al.~\cite{mei2007:Kashibadze_n_en} have taken into account
their own distance estimates. Two cases~(\mbox
{\cite{ton2001:Kashibadze_n_en,tul2008:Kashibadze_n_en}}) derive
the peculiar velocity of the Virgo cluster \mbox{$V_{c,\,{\rm
pec}}=V_c-73 D_c$} from models.

We have estimated $V_c, D_c$ and  $V_{c,\,{\rm pec}}$ values using
the Table~\ref{table01:Kashibadze_n_en} data on radial velocities
and distances of galaxies. The mean values for galaxies being
located inside the virial core of the cluster, as well as for
those residing within the $R_0$ sphere, are presented in the lower
part of the Table~\ref{table04:Kashibadze_n_en}. In doing so we
have considered Virgo members with accurate and supplement
distances separately. The difference between the obtained values
is within the statistical error. Our $V_c$ estimates were
corrected by +40~km~s$^{-1}$ for restriction of velocity range:
$V_{\rm LG} < 2000$~km~s$^{-1}$ with Virgo members velocity
dispersion of 600~km~s$^{-1}$. The following values: $V_c =
984$~km~s$^{-1}$, $D_c = 16.65$~Mpc and  $V_{c,{\rm pec}} =
-231$~km~s$^{-1}$ have been adopted as optimal parameters.

\begin{figure} 
\includegraphics[width=0.48\textwidth]{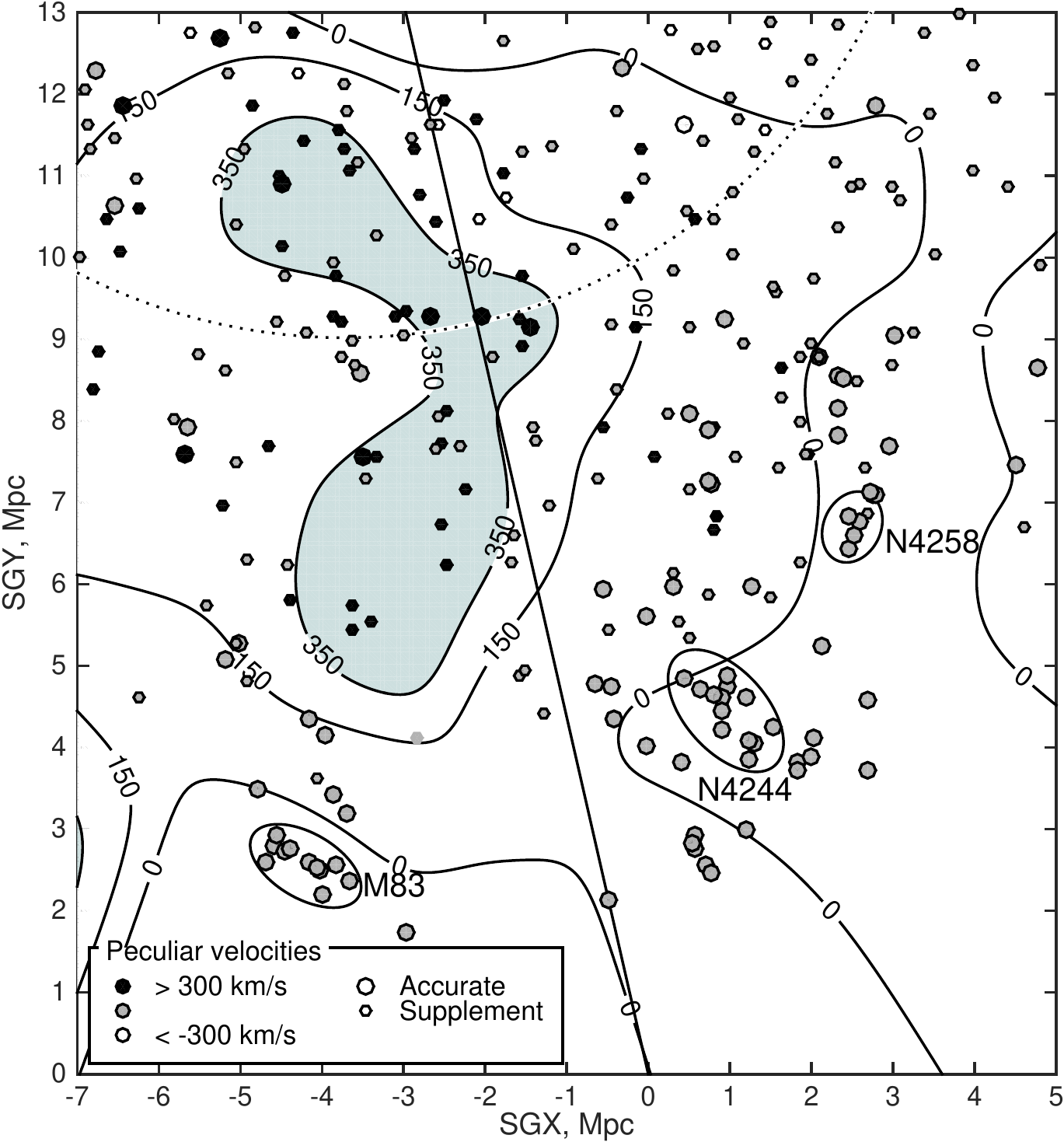}
\caption{Peculiar velocity field at the proximal side of the Virgo cluster.}
\label{fig12:Kashibadze_n_en}
\end{figure}

The number of galaxies with accurate velocities laying behind the
Virgo cluster is small. The linear distance error for background
galaxies with TF distances amounts up to 4--8~Mpc being comparable
with $R_0$. That is why foreground galaxies with accurate
distances remain the main source of data to probe the virgocentric
infall. The distribution of these galaxies projected onto the
Local Supercluster plane in Cartesian supergalactic coordinates is
presented in Fig.~\ref{fig12:Kashibadze_n_en}. The arc of the
circle denotes the zero-velocity surface with radius $R_0$ and
center located at \mbox{${\rm SGX} = -3.7$~Mpc}, ${\rm SGY} =
16.1$~Mpc. The in\-di\-vi\-dual peculiar velocities of galaxies
are smoothed by a Gaussian filter with $\sigma=0.75$~Mpc, and the
resulting peculiar velocity field is outlined by level curves (in
km~s$^{-1}$). Locations of three nearby groups: M\,83, NGC\,4244
and NGC\,4258 are marked by ellipses.

As it can be seen, the zone of maximum positive velocities has an irregular
shape rather elongated in the direction toward the center of the Local
Supercluster. The infall velocity reaches approximately $500$~km~s$^{-1}$. The amplitude
decrease with angular distance from the Virgo center due to the projection
effects being lost among unknown possible variations in tangential motions.

\begin{figure*} 
\includegraphics[width=0.8\textwidth]{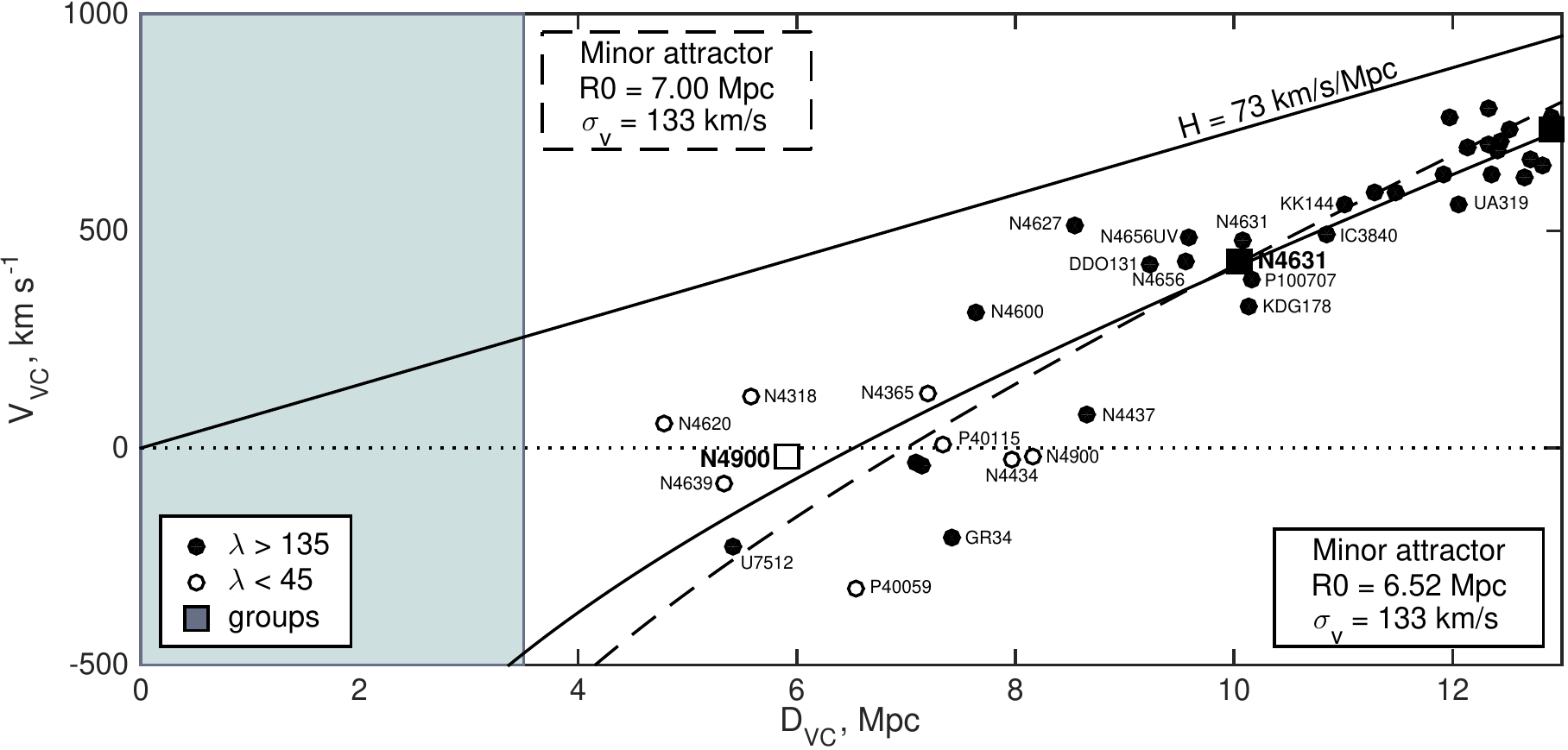}
\includegraphics[width=0.8\textwidth]{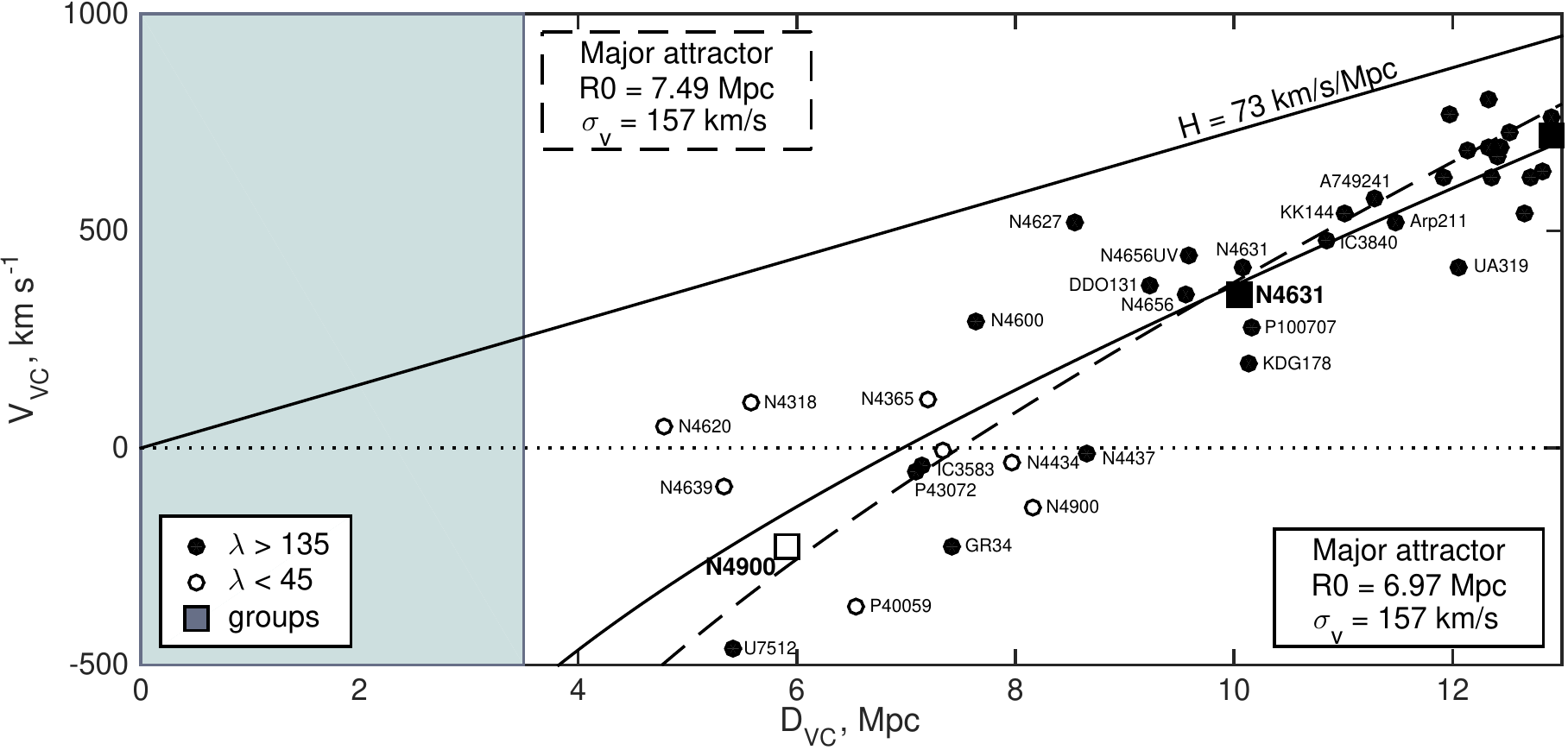}
\caption{Hubble diagram: velocity vs. distance relative to the center
of the Virgo cluster. Top and bottom panel correspond to the minor
attractor and major attractor models. The central zone omitted from
the analysis of peculiar velocities is denoted with grey. Individual
galaxies are marked by circles, galaxy groups---by squares. Solid and
open symbols stay for galaxies located in front and behind the
cluster, respectively.} \label{fig13:Kashibadze_n_en}
\end{figure*}

We have selected galaxies with accurate distances and appropriate
aspect: $\lambda=$[0--45$^{\circ}$] and\linebreak
$\lambda=$[135\mbox{--}180$^{\circ}]$, where $\lambda$ is the
angle between the line of sight and the vector connecting the
cluster center with the position of a galaxy. The Hubble diagram
based on the selected data is shown in
Fig.~\ref{fig13:Kashibadze_n_en}. The top and the bottom panels
correspond to the minor attractor and major attractor models.
Galaxies located in front of the cluster and behind the cluster
are marked with solid and open circles, respectively. Besides
them, we have plotted the centers of groups with accurate mean
distances and $\lambda$ satisfying the condition above as squares.
The central zone $D_{\rm vc}<3.5$~Mpc omitted from the analysis of
peculiar velocities is obscured. The direct/reverse regression
lines for perturbed Hubble flow intersect the zero velocity level
at 6.52 and 7.00~Mpc (minor attractor) and 6.97 and 7.49~Mpc
(major attractor).

We have adopted $R_0=(7.0\pm0.3)$~Mpc as the mean estimate for the
radius of the Virgo zero-velocity surface, and the corresponding
value of $M_T=(7.4\pm0.9)\times10^{14}M_{\odot}$ for the total
mass of the cluster. This value is consistent with virial mass
estimates for the Virgo Cluster:
${(7.5\!\pm\!1.5\!\times\!10^{14}M_\odot}$~\cite{tul1984:Kashibadze_n_en}
and $7.0\times10^{14}M_\odot$~\cite{ton2000:Kashibadze_n_en},
corrected for the Hubble parameter
$H_0=73$~km~s$^{-1}$~Mpc$^{-1}$. The remarkable agreement between
the Virgo mass estimates based on internal (virial) and external
motions of galaxies indicates that the cluster peripheral regions
between $R_V\simeq 2$~Mpc and $R_0\simeq7$~Mpc do not contain a
large amount of dark matter.

\section{CONCLUDING REMARKS}
We have considered a sky region of $100^{\circ}\times 20^{\circ}$
stretched along the Local Supercluster equator and centered at M\,87
being known as the core of the Virgo cluster. This area occupies only
5\% of the whole sky and contains 2158 galaxies with radial
velocities $V_{\rm LG}<2000$~km~s$^{-1}$, which is about 40\% of the
total number of galaxies limited by velocity with 2000~km~s$^{-1}$.
We have determined Tully--Fisher distances $D$ for 567 galaxies from
our sample. Using literature data on distance estimates, the
resulting ratio of LSC belt galaxies with peculiar velocities $V_{\rm
pec}=V_{\rm LG}-73D$ amounts to 52\%. The virial core of the Virgo
cluster and the infall zone are well represented in the considered
region, as well as the vast field areas.

The distribution of early type galaxies (E, S0) shows the well-known
concentration toward the core of the Virgo cluster. The late type
galaxies (S, I, BCD) are mostly presented beyond the virial radius.
Gas-rich dwarf galaxies do not demonstrate any notable concentration
toward the cluster.

Among  50 galaxy groups in the LSC belt, 6 groups have mean
peculiar velocities 500--1000~km~s$^{-1}$ comparable with virial
velocity dispersion of the Virgo galaxies. The cloud consisting of
20--30 galaxies around NGC\,4278 (Coma\,I) is located at about
16~Mpc away from us and about 5~Mpc from the Virgo center, moving
toward an observer with a velocity of 840~km\,s$^{-1}$. The nature
of such an anomaly is not clear and requires the most careful
consideration.

Galaxies in the vicinity of the Virgo cluster are involved in the
virgocentric infall with an amplitude of about $500$~km\,s$^{-1}$.
Based on accurately measured distances, we have estimated the
radius of the zero-velocity surface for the Vurgo cluster to be
\mbox{$R_0=(7.0\pm0.3)$}~Mpc. This value corresponds to the total
mass of the cluster $M_T=(7.4\pm0.9)\times 10^{14}M_{\odot}$ in
good agreement with its virial mass estimate. Such an agreement
indicates that the peripheral regions of the Virgo cluster within
nearly $4R_V$ do not contain a significant amount of dark matter.
The similar result was obtained for the Local Group and other
neighboring groups~\cite{kas2018:Kashibadze_n_en}.

While preparing this paper, we have become aware of the work of Shaya~et~al.~\cite{sha2017:Kashibadze_n_en}
on the dynamics of the Local Supercluster in which the authors give the
Virgo $R_0$ and $M_T$ estimates for the Virgo cluster consistent with our
results within statistical errors.

\acknowledgements The authors are grateful to the Russian Science
Foundation (RSF grant No.~14-12-00965). The authors acknowledge
the usage of the HyperLeda (\url{http://leda.univ-lyon1.fr}) and
NED (\url{http://ned.ipac.caltech.edu}) databases. The authors
thank Dmitry Makarov for his useful comments.

\onecolumngrid

\input{Table1_s_en.tex}
\input{Table2_en.tex}
\input{Table3_en.tex}

\input{Table4_en.tex}

\end{document}

%% file: Table1_s_en.tex
\small
\begin{table}[h]
\clearpage
\caption{Galaxies in the Local Supercluster band} 
\begin{tabular}{lccrlrrllll} \hline\label{table01:Kashibadze_n_en}
 Name                      & RA(J2000.0) Dec&      SGL SGB &  $V_{LG}$& type&  $B_T$&   $W_{50}$& $m_{21}$&    $m-M$&   $D$&    meth\\
\hline
			&	          &     deg    &  km/s   & &    mag  & km/s & mag  &  mag &  Mpc&\\
\hline
  (1)                     &    (2)     &         (3)    &   (4) &(5) &  (6) &  (7) &  (8) &   (9) &  (10) & (11)\\
\hline
UGC05460                 & 100809.2+515040  & 54.42$-$10.53  & 1146 &Sd  & 13.88&   93 & 14.87 & 31.23& 17.80 & tf\\
UGC05459                 & 100810.1+530501 &  53.50$-$09.68  & 1170& Scd & 13.19 & 274 & 13.61 & 31.56 &20.50 & tf\\
PGC2427800               & 101306.5+525646  & 54.11$-$09.23  & 1545& BCD & 17.40    &$-$     &$-$     &$-$      &$-$      & $-$\\
SDSSJ101540.57+523202.4  & 101540.6+523202 &  54.69$-$09.21  & 1206 &I   & 18.15  & 37 & 18.40 & 31.27 &17.99 & TF\\
UGC05571                 & 101942.4+520356 &  55.46$-$09.06  &  724 &Im  & 16.45  & 51 & 15.34 & 29.58&  8.24 & TFb\\
SDSSJ102225.42+475218.3  & 102225.5+475218 &  59.01$-$11.42  & 1650 &Sm  & 18.14    &$-$     &$-$        &$-$      &$-$  &    $-$\\
PGC030715                & 102701.8+561614 &  52.90$-$05.53 &   914& BCD & 16.08  & 58 & 16.24 & 30.18& 10.84 & TF\\
UGC05676                 & 102904.9+544301 &  54.30$-$06.29  & 1499 &Sdm & 14.81 & 123 & 16.05&  31.72 &22.08 & TF\\
PGC2381991               & 103108.9+504709 &  57.60$-$08.49  &  983 &Sm  & 16.58    &$-$     &$-$       &$-$     &$-$  &    $-$\\
NGC3264                  & 103219.6+560502 &  53.51$-$05.08  & 1010& Sd  & 12.61 & 143 & 14.47 & 31.40& 19.43&  tf\\
UGC05720                 & 103231.9+542404 &  54.86$-$06.09  & 1510& BCD & 13.16  &142 & 15.92 & 31.50& 20.00 & TF\\
UGC05740                 & 103445.9+504606 &  57.97$-$08.05  &  698 &Sm  & 15.59 & 138&  14.91&  32.29& 28.73 & TFb\\
PGC2277751               & 103512.1+461412 &  61.68$-$10.71  &  530& BCD & 17.46    &$-$     &$-$        &$-$      &$-$  &    $-$\\
PGC2368186               & 103559.8+501631 &  58.48$-$08.19  &  896& I   & 17.64    &$-$     &$-$ &$-$ &$-$ &$-$\\
PGC2302764               & 103625.0+474153 &  60.62$-$09.67  & 1590 &BCD & 18.08 &  86 & 17.10 & 32.76& 35.6 &  TFb\\
PGC2425292               & 103636.3+525101 &  56.47$-$06.56  & 1043& BCD & 16.08  &124&  16.31 & 32.66& 34.1  &  TF\\
PGC2302994               & 103825.7+474236 &  60.81$-$09.39   &1611& BCD & 16.73 &  69&  16.59 & 31.78 &22.70 & TF\\
NGC3310                  & 103845.9+533012 &  56.14$-$05.91   &1055& Sbc & 11.28 & 190&  13.29    &$-$      &$-$   &   $-$\\
UGC05791                 & 103926.9+475650 &  60.72$-$09.12 &  890 &Sbc & 14.50 & 157 & 15.98 & 32.11& 26.40 & tf\\
UGC05798                 & 103947.1+475557 &  60.76$-$09.08 & 1574& Scd & 14.45 & 184&  15.76 & 32.54& 32.30 & tf\\
SDSSJ103950.97+564402.9  & 103951.3+564401 &  53.63$-$03.85 & 1216& Sm  & 16.90    &$-$     &$-$       &$-$      &$-$  &    $-$\\
PGC2336386              &  104042.4+491224 &  59.81$-$08.21 & 1544 &Im  & 17.10 & 117 & 16.20 & 32.72& 34.9 &  TFb\\
PGC031888               &  104214.2+474600 &  61.14$-$08.84 & 1562& S0  & 15.29    &$-$     &$-$      &$-$      &$-$   &    $-$\\
PGC2362930               & 104251.9+500619 &  59.28$-$07.40 & 1670& BCD & 16.94    &$-$     &$-$      &$-$       &$-$  &    $-$\\
SDSSJ104407.79+474242.1 &  104407.8+474242 &  61.37$-$08.61 & 1570& BCD & 17.63    &$-$     &$-$      &$-$      &$-$   &    $-$\\
UGC05848                &  104423.1+562517 &  54.25$-$03.53 &  902 &Sm &  15.14 & 133&  15.41&  31.98& 24.9  & TF\\
KKH062                  &  104455.7+541225 &  56.11$-$04.76 & 1070& I  &  18.2  &  41 & 17.11 & 30.66& 13.55 & TFb\\
NGC3353                 &  104522.4+555737&   54.71$-$03.69 & 1020& Sb  & 13.22 &  90&  14.94 & 31.42& 19.30&  tf\\
UGC05883                &  104719.3+540216 &  56.45$-$04.57 &  844& Im  & 15.51  & 65&  15.62 & 30.45 &12.30 & TF\\
UGC05888                &  104745.7+560529 &  54.80$-$03.34 & 1319& Im  & 15.19 & 104&  15.34 & 31.59& 20.78 & TF\\
PGC2288707              &  104747.1+465246 &  62.42$-$08.58 & 1521& Sd  & 17.99    &$-$     &$-$      &$-$       &$-$ &    $-$\\
UGC05917                &  104853.8+464315 &  62.66$-$08.51 &  768& Sm   & 15.17 &  98&  15.75&  31.10& 16.60 & TF\\
SDSSJ105107.83+512013.3 &  105107.8+512013 &  59.01$-$05.62  & 862& I    & 18.24   &$-$     &$-$      &$-$      &$-$   &    $-$\\
UGC05953                &  105118.1+443419 &  64.71$-$09.34 & 1819& S0  &  13.34    &$-$     &$-$       &$-$     &$-$  &     $-$\\
PGC2457837               & 105129.6+540205 &  56.80$-$04.07 & 1400& I    & 16.59   &$-$     &$-$       &$-$      &$-$  &   $-$\\
PGC2554441              &   105132.9+570027 & 54.34$-$02.39 & 1950& Scd  & 16.82    &$-$     &$-$       &$-$      &$-$ &    $-$\\
UGC05976                &  105202.8+553604 & 55.54$-$03.12 & 1289& Sdm  & 15.12 &  93 & 15.35    &$-$      &$-$   &$-$  \\ 
UGC05996                & 105257.4+402242 & 68.47$-$11.35 & 1627 &Sd   & 16.44  & 60 & 15.16 & 31.04& 16.14 & TFb\\
UGC05998                & 105308.4+501705 & 60.07$-$05.94&  1437 &Sm   & 14.74 & 173 & 15.82&  32.34& 29.44 & TF\\
UGC06005                & 105327.7+460111 & 63.69$-$08.24 &  748& Sd   & 16.80    &$-$     &$-$       &$-$      &$-$  &   $-$\\
UGC06016                & 105412.8+541714 & 56.81$-$03.60 & 1583& Im   & 16.10  &139 & 14.60 & 32.00& 25.11 & TFb\\
NGC3448                  &105439.4+541818&  56.83$-$03.53 & 1443 &Sb  &  12.41 & 258&  12.85 & 31.33& 18.45 & TF\\
UGC06029                & 105502.3+494334&  60.71$-$05.99&  1418& I   &  14.19 & 153 & 14.78 & 31.85& 23.4  & TF\\
PGC2389897               &105508.2+511119 & 59.48$-$05.18&  1423& BCD &  17.73    &$-$     &$-$      &$-$       &$-$ &    $-$\\
NGC3458                  &105601.5+570701 & 54.58$-$01.83&  1856 &S0  & 13.05    &$-$     &$-$      &$-$      &$-$  &    $-$\\
UGC06041               &  105617.5+413603 & 67.75$-$10.17&  1472& Sd  & 16.95 &  95&  16.54 & 32.57& 32.7  & TFb\\
UGC06039                & 105620.9+564535&  54.91$-$01.99&  1932& Scd & 16.23    &$-$   &   $-$  &$-$  &$-$  &$-$ \\
PGC032915               & 105658.7+500826&  60.53$-$05.51&  1427& BCD & 17.20   &63 & 16.33 & 31.08 &16.44 & TFb\\
PGC2351264              & 105849.6+494258 & 61.05$-$05.48 &  880& BCD & 16.79    &$-$   &  $-$ &$-$ &$-$ &$-$\\
PGC033085               & 105918.9+423523 & 67.19$-$09.18&  1567& BCD & 17.34    &$-$   &  $-$ &$-$ &$-$ &$-$\\
PGC033137               & 110000.2+542532&  57.15$-$02.81 & 1077&  Sdm & 16.79  & 39 & 16.41 & 30.46& 12.34  &TFb\\
\end{tabular}
\end{table}
\clearpage

%% file: Table2_en.tex
\begin{table}
\caption{List of 50 MK-groups within the LSC band}
\begin{tabular}{lrrrrrr} \hline\label{table02:Kashibadze_n_en}
 Group     &  SGL    &    SGB    & $N_v$&  $<V_{LG}>$&     $D$&    $V_{pec}$\\
\hline
	   &  deg     &   deg    &   &  km/s &    Mpc &  km/s\\
\hline
  (1)    &  (2)      &  (3)      & (4) & (5)  &   (6)  &  (7) \\ 
\hline
NGC3458  & 54.59  & $-$1.83 &   6 & 2000 &      $-$  &   $-$\\
NGC3610  & 54.68  &  1.57 &  19 & 1794 &    28.3 & $-$272\\
NGC3665  & 72.93 &  $-$6.81 &  11 & 2038 &    30.8 & $-$210\\
NGC3769  & 65.71  & $-$0.73 &   6 &  780 &    15.1  &$-$322\\
NGC3838  & 56.91 &   4.25 &  11 & 1368  &   29.0 & $-$749\\
NGC3877  & 66.64 &   0.40 &  21 &  955  &   16.4 & $-$242\\
NGC3900  & 85.83 &  $-$6.90 &   4 & 1745 &    31.0 & $-$518\\
NGC3992  & 61.87  &  4.30 &  72 & 1097 &    18.5 & $-$253\\
NGC4062  & 82.41  & $-$2.13 &   4 &  736 &    14.5 & $-$322\\
NGC4111  & 72.11  &  2.22&  20 &  851 &    16.8&  $-$375\\
NGC4123  &110.43  &$-$10.46 &   5 & 1150 &    16.9 & $-$84 \\
NGC4151  & 75.78  &  1.63&  16 & 1031 &    14.7 & $-$ 42\\
NGC4150  & 84.29  & $-$1.32 &   4 &  211 &    16.2&  $-$972\\
NGC4157  & 65.30  &  5.29 &   8 &  834 &    17.5 & $-$443\\
NGC4189  &100.66  & $-$6.01 &   6 & 1987 &    30.0 & $-$203\\
NGC4217  & 68.78  &  4.98 &   5 & 1085 &    19.1 & $-$309\\
NGC4216  &101.08  & $-$5.60 &  16 &   55 &      $-$ &    $-$\\
NGC4244  & 77.73  &  2.41 &   8 &  291&      4.3 & $-$ 23\\
NGC4258  & 68.74 &   5.55 &  15 &  551 &     7.7 & $-$ 11\\
NGC4261  &108.38  & $-$6.94 &  87  & 2060 &    29.4 & $-$ 86\\
NGC4274  & 85.67  &  0.34 &  14   &  990&     16.3 & $-$200\\
NGC4303  &109.86  & $-$6.72 &  23 & 1387 &      $-$  &   $-$\\
NGC4321   &99.03  & $-$3.18 &  17 & 1515 &      $-$  &   $-$\\
NGC4346  & 69.26  &  6.19 &   5 &  787 &    16.4 & $-$410\\
NGC4342  &107.50  & $-$5.57 &   5 &  596 &      $-$ &    $-$\\
NGC4402 & 101.86  & $-$3.24 &   4 &  117&       $-$  &   $-$\\
NGC4472 & 107.02  & $-$3.85 & 355 &  992 &    16.5 & $-$212\\
NGC4490  & 74.79  &  5.94 &   8 &  583 &     8.9 & $-$ 67\\
NGC4527  &112.47  & $-$4.30 &  18&  1592 &    14.5&   534\\
NGC4535 & 107.15  & $-$2.71 &  23 & 1747 &      $-$  &   $-$\\
NGC4546 & 118.80  & $-$5.72 &   4 &  879 &    11.9 &   10\\
NGC4552 & 103.06  & $-$1.17 &  12 &  230 &      $-$  &   $-$\\
NGC4565  & 90.21  &  2.76 &  11 & 1191 &    14.0 &  169\\
NGC4594  &126.69  & $-$6.68 &  11 &  856 &     9.6 &  155\\
NGC4631  & 84.22  &  5.74 &  28 &  635 &     7.4 &   95\\
NGC4636  &113.03  & $-$2.20 &  32 &  757 &    17.8&  $-$542\\
NGC4643 & 113.75  & $-$2.27 &   9 & 1195 &      $-$ &    $-$\\
NGC4666 & 116.23  & $-$2.50 &  16 & 1427 &    16.7 &  208\\
NGC4697 & 121.60  & $-$3.11 &  37 & 1175 &    14.9 &   87\\
NGC4753  &117.43  & $-$0.96 &  23 &  992 &    19.7 & $-$446\\
NGC4808  &112.36  &  1.36 &   5 &  591  &     $-$ &    $-$\\
NGC4856 & 131.23  & $-$3.02  &  5 & 1189 &    19.4 & $-$227\\
NGC4866 & 103.07  &  4.84 &   5 & 1909 &    30.1 & $-$288\\
NGC4900 & 114.42  &  2.03 &   8 &  779 &    21.0 & $-$754\\
NGC4995 & 124.98  &  1.38  &  4 & 1569 &    21.3&    14\\
NGC5054 & 133.93  &  0.61  &  7 & 1556 &      $-$  &   $-$\\
NGC5078 & 144.44  & $-$1.85  & 26 & 1849 &    26.7&  $-$100\\
NGC5084 & 139.13  & $-$0.13 &  12  &1560 &    22.4 & $-$ 75\\
NGC5170  &136.11  &  3.15  &  4 & 1313 &      $-$  &   $-$\\
NGC5236 & 147.93  &  0.99  & 12 &  321 &     4.9 & $-$ 37\\
\hline
\end{tabular}
\end{table}

%% file: Table3_en.tex
\begin{table}							
\caption{Galaxies around NGC~4278 with high peculiar velocities and some other probable companions to NGC~4278}
\begin{tabular}{lcrclrcr} \hline\label{table03:Kashibadze_n_en}
 Name        &  RA (2000.0) Dec & T & $V_{LG}$  &$D$&  meth &$ B_T$&  $V_{pec}$\\
\hline
             &                &     & km/s& Mpc &    &  mag & km/s\\
\hline
  (1)         &        (2)    &  (3)& (4)& (5)&  (6)  &(7) &  (8)\\
\hline
DDO97         & 114857.2+235016&  10 &  452 &13.7&  bs & 15.1 & $-$548\\
UGC6881        &115444.7+200320&  10 &  522 &16.4 & TF & 15.8 & $-$675\\
PGC4561602    & 115504.2+282053&  $-$2 &  458 & $-$   & $-$  & 17.8 &   $-$\\
AGC731823     & 115522.5+282030&   9 &  512 &23.8 & TF & 16.4 &$-$1225\\
KDG82         & 115539.4+313110&   8 &  558 &16.6 & TF & 14.8 & $-$654\\
BTS76          &115844.1+273506&  10 &  407 &15.  &rgb & 16.5 & $-$688 \\ 
KUG1157+31     &120016.2+311330&   8 &  582& 14.8 & TF&  15.1 & $-$498\\
NGC4080       & 120451.8+265933&   8 & 517 &15.0 & TF & 13.7 & $-$578\\
LV1205+28     & 120534.2+281355&   8 & 462 &19.5 & TF & 16.7 & $-$962\\
PGC4560429    & 120633.5+303716&  $-$2 & 464 & $-$   & $-$  & 18.2 &   $-$\\
UGC7131       & 120911.8+305424 &  8 & 224 &16.8 & TF & 15.7 &$-$1002\\
NGC4150       & 121033.6+302406&  $-$1 & 210 &13.7 &sbf & 12.5&  $-$790\\
KK127         & 121322.7+295518&   9 & 103& 17.3 & TF & 15.6& $-$1160\\
AGC732009     & 121348.4+295731&  10  &196 &16.5 & TF & 17.4& $-$1008\\
LV1217+32     & 121732.0+323157 &  9 & 433 &15. & rgb & 18.4&  $-$662\\
UGC7321       & 121734.0+223225&   7 & 339& 17.2&  TF & 14.1&  $-$917\\
AGC229053     & 121815.5+253406&  10 & 376 &17.9&  TF&  17.9&  $-$931\\
KKH07J1218+30 & 121831.5+300340&  10 & 600 & $-$  &  $-$  & 17.0 &   $-$\\
BTS116        & 121857.3+283311&  $-$3 & 289 & $-$  &  $-$ &  16.2 &   $-$ \\
IC 779        & 121938.7+295300&  $-$1 & 187 &16.7& sbf & 15.2& $-$1032\\
PGC0213976    & 121943.6+293932&  $-$2 & 549 & $-$  &  $-$  & 17.3 &   $-$\\
NGC4278       & 122006.8+291651&  $-$2 & 588 &16.1& sbf & 11.1&  $-$587\\
NGC4286       & 122042.1+292045&   2 & 611 &14.7&  TF & 14.5&  $-$448\\
PGC1853813    & 122116.6+290221&  $-$2 & 588 & $-$  &  $-$  & 16.5&    $-$\\
NGC4308       & 122156.9+300427&  $-$1 & 612 & $-$  &  $-$  & 14.1&    $-$\\
PGC4323538    & 122216.7+305324&  $-$1 & 545 & $-$  &  $-$  & 17.1&    $-$\\
UGC07438      & 122219.5+300341&   6 & 678 &22.3&  TF & 15.8 & $-$950\\
PGC040195     & 122309.7+292059&  $-$2 & 524 & $-$   & $-$  & 15.6&    $-$\\
IC 3247       & 122314.0+285338 &  7 & 546 &24.4 & TF & 15.4& $-$1235\\
AGC749235     & 122409.9+261352&  10 & 246&  $-$  &  $-$  & 19.8&    $-$\\
IC 3308       & 122517.9+264253&   8 & 277 &12.8&  TF & 15.4 & $-$657\\
AGC749236     & 122542.4+264836&  10 & 234& 15.1& TFb & 16.6&  $-$868\\
NGC4414       & 122627.1+311324&   5 & 719 &17.9& cep & 11.0&  $-$588\\
NGC4448       & 122815.5+283713&   2 & 659 &23.8&  tf & 12.0& $-$1078\\
UGC7699       & 123248.0+373718&   7 & 514 &14.5 & TF & 13.3&  $-$544\\
UGC7774       & 123622.5+400019&   7 & 563 &22.6 & TF & 14.6& $-$1087\\
FGC1497       & 124700.6+323905&   8 & 522& 23.4 & TF & 16.8& $-$1186\\
UGC7990       & 125027.2+282110&  10 & 495& 20.4 & TF & 16.2&  $-$994\\
\hline
\end{tabular}
\end{table}

%% file: Table4_en.tex
\begin{table}
\caption{The mean radial velocity and mean distance of Virgo cluster}
\begin{tabular}{lrcccl}\hline\label{table04:Kashibadze_n_en}
  Sample           & Number     &   $V_c(LG)$&     $D_c$   &   $V_{c,pec}$ &  Reference \\
                    &          &    km/s   &     Mpc    &   km/s&\\
\hline
 All members  (V)   &  385     &  956$\pm$ 55 &   [16.5]   &   --248&   Binggeli et al, 1993\\
 Core members (V)   &  271     &  935$\pm$ 35 &   [16.5]   &   --270 &  Binggeli et al, 1993\\
 All members  (D)   &   79     & 1034$\pm$ 61 &  16.5$\pm$0.1 &   --170 &  Mei et al, 2007\\
 Core members (D)   &   32     &  984$\pm$105 &  16.7$\pm$0.2 &   --235 &  Mei et al, 2007    \\     
 Infall pattern     &  189     &   --       &     --      &   --139 &  Tonry et al, 2000\\
 Cosmic flow        &    --      &  --       &     --      &   --185 &  Tully et al, 2008\\
\hline
 Within $R_v$, accur   &  75     &  950$\pm$62  &  16.45$\pm$0.08 & --251 &  present paper\\
 Within  $R_0 $, accur   & 119     &  975$\pm$46 &   16.32$\pm$0.16 & --216 &  present paper\\
 Within  $R_0 $, suppl   & 372      & 996$\pm$27  &  16.44$\pm$0.16 & --204 &  present paper\\
 Within  $R_0 $, all     & 491     &  985$\pm$23  &  16.38$\pm$0.14 & --211 &  present paper\\
\hline
 Adopted             &         &  984      &  16.65       & -231 &\\
\hline
\end{tabular}
\end{table}